\pdfoutput=1
\documentclass[12pt,titlepage]{article}
\usepackage[utf8]{inputenc}
\usepackage{aaai}
\usepackage{times}
\usepackage{helvet}
\usepackage{natbib}
\usepackage{courier}
\usepackage{caption}
\usepackage{color}
\usepackage{xcolor}
\usepackage{graphicx}
\usepackage{booktabs}
\usepackage{placeins}
\usepackage[most]{tcolorbox}
\usepackage{titlesec}
\usepackage{balance}
\usepackage[hidelinks]{hyperref}
\usepackage{csquotes}
\usepackage{enumitem}
\usepackage{hyperref}
\usepackage{amsmath}
\usepackage[yyyymmdd,hhmmss]{datetime}

\newcommand{\eg}[2]{
    \begin{tcolorbox}[colback=black!5!white,colframe=black,title={#1},breakable, enhanced]
        #2
    \end{tcolorbox}
}

\newcommand{\oldtext}[1]{\textcolor{red}{#1}}

\begin{document}

\setlength{\footskip}{10cm}
\title{TED-On: A Total Error Framework for Digital Traces of Human Behavior on Online Platforms\\
\large [A condensed version of this article is set to appear in the Public Opinion Quarterly Journal]}

\author{Indira Sen\\
{GESIS}\\
Indira.Sen@gesis.org
\And
Fabian Fl{\"o}ck\\
{GESIS}\\
Fabian.Floeck@gesis.org
\And
Katrin Weller\\
{GESIS}\\
Katrin.Weller@gesis.org
\And
Bernd Weiß\\
{GESIS}\\
bernd.weiss@gesis.org
\And
Claudia Wagner\\
{GESIS \& University Koblenz}\\
Claudia.Wagner@gesis.org}
\maketitle

\pagenumbering{arabic}
\pagestyle{plain}
\begin{abstract}

Peoples’ activities and opinions recorded as digital traces online, especially on social media and other web-based platforms, offer increasingly informative pictures of the public. They promise to allow inferences about populations beyond the users of the platforms on which the traces are recorded, representing real potential for the Social Sciences and a complement to survey-based research. But the use of digital traces brings its own complexities and new error sources to the research enterprise. Recently, researchers have begun to discuss the errors that can occur when digital traces are used to learn about humans and social phenomena. This article synthesizes this discussion and proposes a systematic way to categorize potential errors, inspired by the Total Survey Error (TSE) Framework developed for survey methodology. We introduce a conceptual framework to diagnose, understand, and document errors that may occur in studies based on such digital traces. While there are clear parallels to the well-known error sources in the TSE framework, the new “Total Error Framework for Digital Traces of Human Behavior on Online Platforms” (TED-On) identifies several types of error that are specific to the use of digital traces. By providing a standard vocabulary to describe these errors, the proposed framework is intended to advance communication and research concerning the use of digital traces in scientific social research. 
\end{abstract}

%\keywords{Fake social engagement, Social networks, Instagram}

%% I need this sometimes to remind me of the structure of the text; will be removed later (BW)
\begin{comment}
\begin{tiny}
\tableofcontents
 \end{tiny}
\end{comment}

\section{Introduction}\label{sec:intro}

%Measuring a hypothesized construct and inferring its manifestation in a larger unseen target population from a sub-sample has been at the heart of empirical Social Science studies since the beginning of the discipline, using the gradually refined tools of survey design and probabilistic sampling in particular. Survey methodology has distilled its history of research dedicated to identifying and analyzing the various errors that occur in the statistical measurement of collective behavior and attitudes into the Total Survey Error framework (TSE), which provides a conceptual structure to identify, describe and quantify the errors of survey estimates
When investigating social phenomena, for decades, the empirical social sciences have relied on surveying samples of individuals taken from well-defined populations as one of their main data sources, e.g., general national populations. An accompanying development was the constant improvement of methods as well as statistical tools to collect and analyze survey data~\citep{groves_three_2011}. These days, survey methodology can be considered an academic discipline of its own~\citep{wolf_survey_2016}. It has distilled its history of research dedicated to identifying and analyzing the various errors that occur in the statistical measurement of collective behavior and attitudes as well as generalizing to larger populations into the Total Survey Error framework (TSE). This framework provides a conceptual structure to identify, describe, and quantify the errors of survey estimates~\citep{biemer2010total,groves2010total,weisberg2009total,groves2011survey}. While not existing in one single canonical form, the tenets of the TSE are stable and provide survey designers with a guideline for balancing cost and efficacy of a potential survey and, not least, a common vocabulary to identify error sources in their research design from sampling to inference. 
In the remainder, we will refer to the concepts of the TSE as put forth by~\citet[48]{groves2011survey}.%\footnote{Groves et al. refer to the TSE in their book as ``Survey lifecycle from a design perspective''.} %Bernd: they also speak of "...from a quality perspective". not sure, though, that this footnote is important

\begin{figure}[t!]
\begin{center}
\includegraphics[scale=0.3]{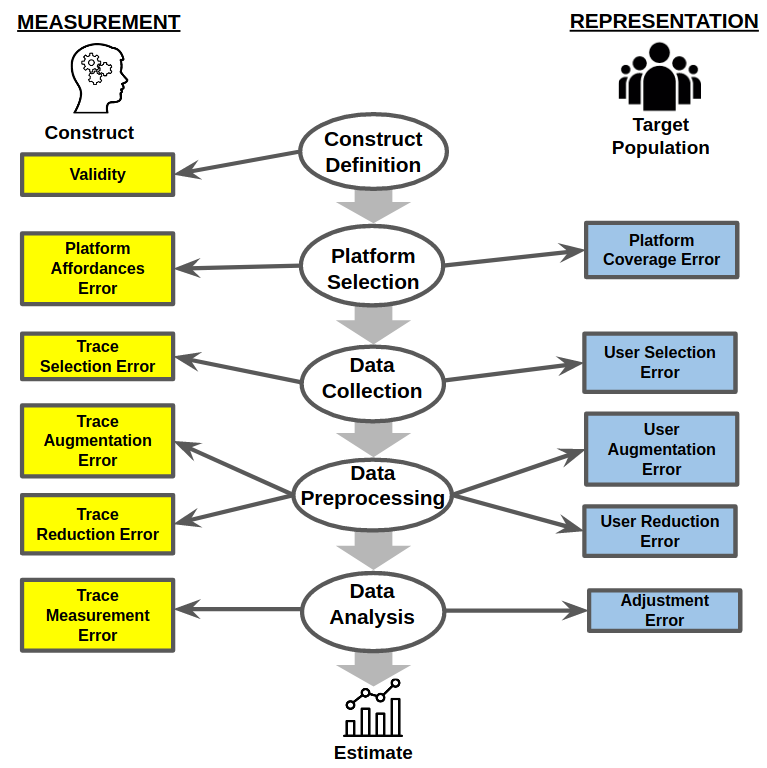}%measurement_representation.pdf}
\end{center}
 \caption{\textbf{Potential measurement and representation errors in a digital trace based study lifecycle.} Errors are classified according to their sources: errors of measurement (due to how the construct is measured) and errors of representation (due to generalising from the platform population to the target population of interest). Errors follow from the design decisions made by the researcher (see steps depicted in the center strand). `Trace' refers to user-generated content and interactions with content (e.g. liking, viewing) and between users (e.g. following), while 'user' stands for the representation of a target audience member on a platform, e.g., a social media account [Best viewed in color]\protect}
 \vspace{-1.5em}
\label{fig:data_coll}
\end{figure}

\footnotetext{Icons used in this image have been designed by \href{https://www.flaticon.com/authors/becris}{Becris}, \href{https://www.flaticon.com/authors/elias-bikbulatov}{Elias Bikbulatov} and \href{https://www.flaticon.com/authors/pixel-perfect}{Pixel perfect} from \url{https://www.flaticon.com}.}

Recently, surveys have come to face various challenges, such as declining participation rates, while simultaneously, there has been a growth of alternative modes of data collection~\citep{groves_three_2011}. In the course of continuous digitalization, this often includes new types of data that have not been collected in a scientifically pre-designed process but are captured by web-based platforms and other digital technologies such as smartphones, fitness devices, RFID mass transit cards or credit cards. These digital traces of human behavior become increasingly available to researchers.
It is especially the often easily accessible data from social media and other platforms on the World Wide Web that has become of heightened interest to scientists in various fields aiming to explain or predict human behavior~\citep{watts2007twenty, lazer2009social, salganik2017bit}.\footnote{Researchers refer to digital trace data with different names including social data~\citep{olteanu2016social,alipourfard2018can} and big data~\citep{salganik2017bit,pasek2019s}.}
Beside studying platforms \textit{per se} to understand user behavior on and societal implications of, e.g., a specific web site~\citep[see ][]{DiMaggio2011}, digital trace data promises inferences to broad target populations similar to surveys, but at a lower cost, with larger samples~\citep{salganik2017bit}. In fact, research based on digital trace data is frequently referred to as ``Social Sensing'' -- i.e., studies that repurpose individual users of technology and their traces as sensors for larger patterns of behavior or attitudes in a population~\citep{an2015whom,sakaki2010earthquake}. 
In addition to increasing scale and decreasing costs, digital traces also capture quasi-immediate reactions to current events (e.g., natural disasters) which makes them particularly interesting for studying reactions to unforeseeable events which surveys can only ask about in retrospect. %Therefore, there is a growing interest in the use of digital traces for mining insights and, predicting present (`nowcasting') and future trends.%, and they are created independently of researchers' requests thus offer unobtrusive approaches for learning about public opinions. 

However, digital traces come with various challenges such as bias due to self-selection, platform affordances, data recording and sharing practices, heterogeneity, size, etc., raising epistemological concerns~\cite{tufekci2014big,olteanu2016characterizing,ruths2014social,schober2016social}. Another major hurdle is created by uncertainty about exactly how the users of a platform differ from members of a population to which researchers wish to generalize, which can change over time. While not all of these issues can be mitigated, they can be documented and examined for each particular study that leverages digital traces, to understand issues of reliability and validity (e.g., \citet{lazer2015issues}). Only by developing a thorough understanding of the limitations of a study can we make it comparable with others. Moreover, assessing the epistemic limitations of digital trace data studies can often help illuminate ethical concerns in the use of these data (e.g.,~\cite{olteanu2016characterizing,mittelstadt2016ethics,jacobs2019measurement}).

\begin{figure}[t!]
\begin{center}
\includegraphics[scale=0.365]{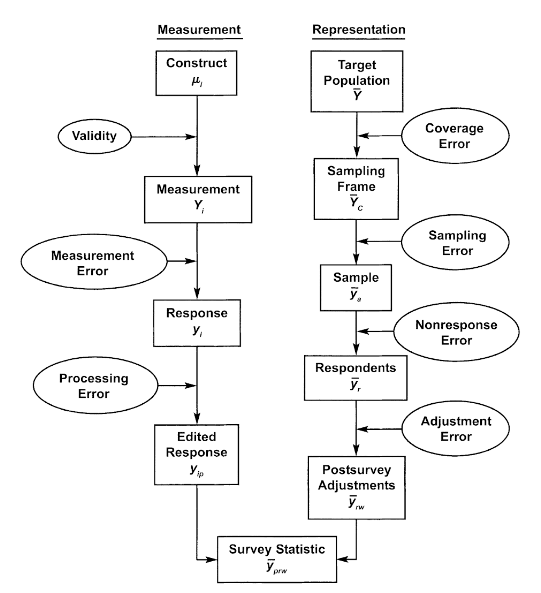}
\end{center}
 \caption{\textbf{Total Survey Error Components} linked to steps in the
measurement and representational inference process~\citep{groves2011survey}}
\vspace{-2em}
\label{fig:tse}
\end{figure}

%\textbf{Our Contributions.} In this work, we put forth a framework that allows to describe, analyze and mitigate errors that occur in digital trace based studies that aim to make inferences about a theoretical construct (see Figure \ref{fig:data_coll}). We describe the relations between our framework and the Total Survey Error Framework to bridge the vocabularies of different communities.
%Our framework is mainly inspired by -- but not limited to -- social media data such as Twitter, Facebook, and Reddit data as well as other data collected on web platforms such as search engine queries, used in ''Social Sensing'' studies.
%\footnote{Choi et al, provide instances where search engine data was used to forecast various economic indicators~\citep{choi2012predicting}}. %Social Media (SM) data, concentrated on -- but not limited to -- the examples of Twitter, Facebook, Reddit and similarly structured SM platforms that have received the majority of attention in Social Sensing studies. 

%Concretely, the goal of our proposed framework is to describe errors that can occur in studies that (i) use digital traces in the form of (textual or audiovisual) content produced by users and other (inter)actions of users on specific platforms to (ii) make inferences about a theoretical construct in a larger target population of humans, beyond those platforms.

%\footnote{In Section \ref{}, we discuss how the framework can also be applied to intra-platform studies and to studies with analysis objects other than humans.}

\textbf{Our Contributions.} Our work adds to the growing literature on identifying errors in digital trace based research~\citep{olteanu2016social,tufekci2014big,hsieh2017total,ruths2014social,lazer2015issues}, by highlighting these errors through the lens of survey methodology and leveraging its systematic approach. Based on the TSE perspective, we propose a framework that covers the known error sources potentially involved when using digital trace data, the Total Error Framework for Digital Traces of Human Behavior on Online Platforms (TED-On). This allows researchers to characterize and analyze the errors that occur when using data from online platforms to make inferences about a theoretical construct (see Figure 1) in a larger target population beyond the platforms’ users. By connecting errors in digital trace-based studies and the TSE framework, we establish a common vocabulary for social scientists and computational social scientists to help them document, communicate and compare their research. The TED-On, moreover, aims to foster critical reflection on study designs based on this shared vocabulary, and consequently better documentation standards for describing design decisions. Doing so helps lay the foundation for accurate estimates from web and social media data. In our framework we map errors to their counterparts in the TSE framework, and, going beyond previous approaches which leverage the TSE perspective~\citep{amaya2020total,japec2015big}, describe new types of errors that arise specifically from the idiosyncrasies of digital traces online, and associated methods. Further, we adopt the clear distinction between measurement and representation errors for our framework (cf. Figure 1), as proposed by~\cite{groves2011survey}. Through running examples (and a case study in Supplementary Materials, Section 3) that involve different online platforms including Twitter and Wikipedia, we demonstrate how errors at every step can, in principle, be discovered and characterized when working with web and social media data. This comprises measurement from heterogeneous and unstructured sources common to digital trace data, and particularly the challenge of generalizing beyond online platforms. 
While our framework is mainly focused on web and social media data, it is in principle applicable to other sources of digital trace data of humans as well, such as mobility sensors, transaction data or cell phone records. We discuss such further applications at the end of this paper.

\section{Background: The Total Survey Error and its Relevance to Digital Trace Data}\label{sec:tse}

% \begin{figure*}[!htb]
% \begin{center}
% \includegraphics[scale=0.7]{figs/svd.pdf}
% \end{center}
% \caption{\textbf{The divergence between survey estimates and digital trace data estimates}. The primary point of difference is that digital trace data is entirely nonreactive due to the lack of a solicitation of (often standardized) responses. In surveys, respondents are surveyed, and their responses are used to construct the final estimate, whereas, in digital traces, a subset of users (which are usually users) of a platform or their traces are used to construct the estimate. The captions in the pink boxes refer to the relevant stage of a digital trace based study lifecycle -- they are not necessarily correlated with the steps in the survey pipeline above.% Dashed arrows indicated the exclusion of relevant steps in this ideal process that, however, are not important for this comparison.
% \protect\footnotemark}
% %\small\textsuperscript{} Icons used in this image have been designed by [author link] from www.flaticon.com
% \label{fig:survey_digital}
% \end{figure*}

In this section, we explain the differences and commonalities between a survey-based and a digital trace data study of human behavior or attitudes. We do so to develop an understanding of the limitations and strengths of both approaches to conduct social science research and, thus, the sources and types of errors that affect both approaches. We also explore the Total Survey Error Framework and how it helps survey designers document their studies.

While both processes have some overlapping stages (see section~\ref{sec:digital_trace_research}), the primary point of difference is that digital trace data is typically nonreactive due to the lack of a solicitation of (often standardized) responses. %Figure~\ref{fig:survey_digital} illustrates this difference between survey-based and digital trace data studies. 
In survey-based research, careful planning is required to come up with questions that measure the construct of interest and a sensible sampling frame from which a (random) sample of individuals, households or other units is drawn. Next, an interviewer- or self-administered survey is conducted and the responses are collected, which are then used to estimate the population parameter. In this research design, though, respondents are well aware that they are subject to a scientific study, entailing several potential consequences, such as social desirability bias in answers. And, at each stage of the survey life cycle, data quality can be compromised due to a multitude of errors (for more information, see section \ref{sec:tse-framework}). 
When relying on digital trace data collected via a web platform, there is no need for developing a sampling design or interviewing respondents. Here, individuals' traces (or a subset of traces after querying) are readily available. Hence, individuals are usually not aware that they are subject to a scientific investigation, which has the advantage that researchers can directly observe how users behave rather than relying on \textit{self-reported } data. It has the disadvantage that researchers may run the risk of misunderstanding these traces, as they are produced in complex and heterogeneous technical and social contexts. This can be avoided in a survey by constructing effective questionnaires and controlling the survey situation to some extent. Much of the divergence between surveys and digital trace based study rests on how we may effectively use potentially noisy, unsolicited traces to understand theoretical constructs while keeping in mind that these traces may not be effective proxies for that construct. In the next subsection, we describe a typical survey workflow through the lens of the TSE.

\footnotetext{Icons used in this image have been designed by \href{https://www.flaticon.com/authors/hadrien}{Hadrien}, \href{https://www.flaticon.com/authors/becris}{Becris}, \href{https://www.flaticon.com/authors/freepik}{Freepik}, \href{https://www.flaticon.com/authors/smartline}{Smartline}, \href{https://www.flaticon.com/authors/pixel-perfect}{Pixel perfect}, \href{https://www.flaticon.com/authors/pixelmeetup}{pixelmeetup}, \href{https://www.flaticon.com/authors/eucalyp}{eucalyp}, \href{https://www.flaticon.com/authors/gregor-cresnar}{Gregor Cresnar}, and \href{https://www.flaticon.com/authors/prettycons}{prettycons} from \url{www.flaticon.com}}

\subsection{The Total Survey Error Framework}\label{sec:tse-framework}

In survey research, errors can be described and systematized by focusing on the \textit{source} of errors, most prominently the distinction between measurement errors and representation errors \citep{groves2011survey}.\footnote{We refer to error as the difference between the obtained value and the true value we want to measure, while bias refers to systematic errors, following the definitions in Weisberg's `The Total Survey Error Approach', p.22~\citep{weisberg2009total}} For our framework, we mostly focus on Groves et al.'s approach in making an overall distinction between errors in, firstly, defining and measuring a theoretical construct with chosen indicators \textit{(measurement errors}) and, secondly, the errors arising when inferring from a sample to the target population (\textit{representation errors}) as illustrated in Figure \ref{fig:tse}~\citep{groves2011survey}. We adopt this specific distinction as it is helpful in conceptually untangling different fallacies plaguing surveys as well as related research designs with non-designed, non-reactive, and non-probabilistic digital data. %We hope to explore the repercussions of incorrect handling of units of analysis in the future. 
The reader should keep in mind that errors can be both systematic (biased) or randomly varying at every step of the inferential process we describe in the remainder.

\textbf{Measurement errors}, from a survey methodological perspective, can be regarded as the ``[...] departure from the true value of the measurement as applied to a sample unit and the value provided" \citep[52]{groves2011survey}.
%\footnote{Groves' TSE also identifies ``edited response'' or ``processing errors'' as measurement errors \citep{groves2010total, groves2011survey}, which are not core problems of construct validity. \citet{straub2004validation} identify content validity as a separate issue, which we subsume under construct validity for simplicity here, following \citet{guion1977content} and \citet{rogers1995psychological}.} 
With respect to the \enquote{Measurement} arm in Figure~\ref{fig:tse} representing those errors, the first step of a survey-driven -- and any similar study -- requires defining the theoretical construct of interest and establishing a theoretical link between the captured data and the construct~\citep{Howison2011}. 
% 
% add some refs from Jungherr paper e.g., Carmines & Zeller, 1979; Cronbach & Meehl, 1955; Kane, 1992; Messick, 1989
%
%Construct validity refers to whether an indicator (i.e., scale, a test or a metric) is an effective measure of the theoretical construct. Construct validity must be considered in any stage of digital trace based study and we will explain in the later section how different biases may impact construct validity at any stage.
%
Survey researchers usually start by defining the main construct of interest (e.g., ``political leaning'', ``personality'') and potentially related constructs (e.g., ``political interest''). This is followed by the development or reuse of scales (i.e., sets of questions and items) able to \textit{measure the construct adequately}, establishing \textit{validity}. In developing scales, content validity, convergent construct validity, discriminant construct validity, internal consistency as well as other quality marks are checked (cf.~\citet{straub2004validation}). The design of items usually follows a generally fixed and pre-defined research question and theorized constructs (notwithstanding some adaptations after pre-testing), followed by fielding the actual survey.~\citet{groves2011survey} further point out ``measurement error'' (not to be confused with the measurement error
arm of the TSE), which we denominate \textit{response error} here for clarity. It arises during the solicitation of actual information in the field even when an ideal measurement has been found. I.e., a respondent understands a set of items as intended, but either cannot (usually trough recall problems) or does not want to (e.g. social desirability) answer truthfully. Lastly arise \textit{processing errors} introduced when processing data, such as coding textual answers into quantitative indicators and data cleaning. Besides validity, individual responses may also suffer from variability over time or between participants, contributing to low \textit{reliability}. 

\textbf{Representation errors} are the second source of error which are concerned with the question of how well a survey estimate generalizes to the target population. Representation errors consist of coverage error, sampling error, nonresponse error and adjustment error \citep{groves2011survey} and contribute to the \textit{external validity} or generalizability of a study. 

Survey design begins its quest for unbiased representation by clearly defining the target population that the construct of interest should be measured for/inferred to, e.g., the national population of a nation-state. Then, a sampling frame is defined, the best approximation of all units in the target population, e.g., telephone lists or (imperfect) population registers, resulting in under- or over-coverage of population elements, constituting \textit{coverage error}. Ineligible units might add noise to the sampling frame, such as business telephone numbers. Note that coverage error ``exists before the sample is drawn and thus is not a problem arising because we do a sample survey''~\citep{groves2011survey}. When actually drawing a (random) sample employing the previously defined sampling frame, \enquote{one error is deliberately introduced into sample survey statistics}, as~\citet[56]{groves2011survey} put it. Mostly due to financial constraints, but also due to logistical infeasibilities, it is not possible to survey every element in the sampling frame. By ignoring these elements, the sample statistics will most likely deviate from the (unobservable) sampling frame statistics, thereby introducing a \emph{sampling error}. The sampling error can be decomposed into two components: sampling variance, the random part, and sampling bias, the systematic part. The sampling variance is a consequence of randomly drawing a set of \(n\) elements from a sampling frame of size \(N\). Applying a simple random sample design, \(\binom{N}{n}\) samples can be realized. Since each sample realization is composed of a unique composition of elements, sampling variance is introduced. Sampling bias as the second component of sampling error comes into play when the sampling process is designed and/or executed in such a way that a subset of units is selected from the sampling frame but giving some members of the frame systematically lower chances to be chosen than others. Of course, sampling error can only arise when there is no feasible way of reaching all elements in the sampling frame -- if one could access all elements of the complete sampling frame with minimal cost, sampling error would not occur. 

Further, if chosen individuals drawn as part of the sample refuse to answer the whole survey, we speak of unit \textit{nonresponse errors}.
%, or item nonresponse %<-- NOTE(FFL): according to Jaeckle, this is measurement error. And I think I would agree. 
%errors if only certain items are not collected from certain respondents (e.g., if they break off the survey process after some questions). 
While in most cases providing insufficient responses to items hinders valid inferences regarding the topical research questions, nonresponse to demographic items %metadata (such as sociodemographic information) 
can also hinder post-survey adjustment of representation errors.\footnote{While not mentioned explicitly by Groves, this affects the adjustment step and becomes much more important when working with digital traces.} Lastly,~\citet{groves2011survey} list \textit{adjustment error}, occurring when reweighting is applied post-survey to under- or over-represented cases due to any of the representation errors described above. The reweighting is usually based on socio-demographic attributes of individuals and often their belonging to a certain stratum.

\subsection{Research with Digital Trace Data}\label{sec:digital_trace_research}

While digital traces are often used to measure the same things as surveys do, they proceed differently
%, as summarized in Figure \ref{fig:survey_digital}
. Like in a survey, the researcher \textbf{defines the theoretical construct and the ideal measurement} that will quantify it. However, in survey-based research, the measurement instrument can be constructed tailored to the research question, including a stimulus (question). In contrast, the researcher has less control over digital-trace-based measurement instruments that need to work with already existing traces, the stimuli for which are not researcher-administered. The researcher then picks the source of the digital traces by \textbf{selecting one or more platforms} (often, on the Web). The chosen platform(s) act(s) as the sampling frame. Depending on data accessibility, all ``users'' on the platform, as well as the ``traces''
produced by them, may be available to the researcher. These together constitute the digital traces of interest. 
\textit{Traces} in our definition refer to (i) user-generated content ~\citep{johnson2015emergence}, containing rich -- but often noisy -- information in form of textual and visual content~\citep{schober2016social} (e.g., tweets, photos, details in their profile biography); and to (ii) records of user activity which provide signals about users who do not post content, but interact with existing posts and other content, e.g., by ``liking'', viewing or downloading. 

Traces are in general generated by identifiable \textit{users} that can be reasonably believed to represent a human actor (or sometimes, a group of human actors). Such users are most frequently platform accounts or user profiles, but might also be IP addresses. In alternative settings, they might be bank accounts, smart devices or other proxies for human actors. Users representing organisations or automated programs run the risk of being mistaken as suitable proxies for human actors and might have to be filtered out (cf. Section \ref{sec:data_preprocessing}). Users typically emit various traces over time and carry additional attribute information about themselves (gender or location of the account holder, type of account, etc.). 

Note that while \textit{traces} are produced by \textit{users}, these concepts are not mappings of \textit{responses} and \textit{respondents} to digital trace data. 
Instead, traces are produced without researcher-induced stimuli, and users do neither always stand for a single element of the (human) target population nor are they always observable. 
In fact, traces cannot always be linked to a user account or another reasonably persistent proxy of a human actor, e.g., in systems that only link traces to dynamically assigned IPs or other quasi-anonymous identifiers (such as Wikipedia pageviews, or single online shop purchases without links to accounts).\footnote{Although these data often exist, they may not be made available to researchers, e.g. due to data protection concerns or commercial interests of platform providers.}
Further, the link from traces and users to the units of the target population -- in order to make inferences about them -- can be constructed in different ways. 
%\textit{or} separate traces (like posts, tweets, likes, etc.). 
This at times limits the understanding of what the eventual aggregate statistic represents in digital trace studies, since it might be (a) an aggregate of traces collected inside a particular boundary without specifying a concrete set of users to be collected (e.g., all traces in a certain time period or at a certain location), or (b) take into account these users, e.g., by aggregating social media posts per user profile and reweighting them by attached socio-demographical metadata. If only traces are used, they cannot be assumed to be independent traces from an equal amount of human ``emitters'' or ``sensors''. This can lead to methodological as well as epistemological issues. 
%We want to highlight, in fact, that for any study with non-reactive, non-probabilistic digital trace data that aims to make inferences about a target population of humans, what is regarded as a proxy for an element of the target population always should be these users, not single traces of those users. 

For \textbf{data collection}, and in contrast to surveys, researchers can then (i) chose to either sample traces \textit{or} users, and (ii) frequently have access to the entire sampling frame of a platform (in this case, one might even speak of a census of the platform). Therefore, there is often no need for sampling due to logistical infeasibility, as is typically done in surveys - and ``sample'' has to be understood in this light. Though the entire sampling frame %(or census, if the platform is identical to the study population)
is technically collectable, the researcher often still needs to create a subset of traces or users since it is no necessarily storable or processable because of its volume, due to restrictions imposed by data providers\footnote{The best-known example: Twitter's restricted-access streaming APIs which enforce a platform-provided ``random sample'' as described by Twitter.} or other legal or ethical restrictions. Depending on their needs and the construct being studied, the researcher thus devises a data collection strategy, which usually involves a set of \textit{queries} that select a specific portion the platform data, producing the final data subset that will be used for the study.\footnote{We use the term `platform data' to refer to users or traces generated by users on that particular platform for the rest of the paper.} 

The traces and the users that make up this subset are usually further filtered or \textbf{preprocessed}, including labeling (or coding). In this regard, digital trace data is similar to survey data based on open-ended questions, which also requires a considerable amount of preprocessing before it can be statistically analyzed. However, due to commonly very high volume of digital traces, preprocessing of these data is almost exclusively done via automated methods that rely on heuristics or (semi-)supervised machine learning models. Depending on the subject of analysis, data points are then aggregated and \textbf{analysed}, to produce the final estimate, as is done for survey estimates. 

%For digital trace studies, it is moreover not always explicitly clear what the unit of analysis is, e.g., individual humans, or groups of humans, or all humans in a nation state. 

Note that TED-On is centered around an idealized and somewhat prototypical abstraction of the research process for digital traces. As there is no unified convention for how to conduct such research -- potentially involving a wide range of data and methods -- the actual workflows may differ from this abstraction, e.g., depending on disciplinary backgrounds. 
E.g., in an ideal world, the research design pipeline as outlined in Figure \ref{fig:data_coll} % and \ref{fig:survey_digital} 
would be followed sequentially, whereas in reality, researchers will start their process at different points in the pipeline, e.g., get inspiration in form of an early-stage research question through a platform's available data and its structure, and then iteratively refine the several steps of their research process. This is a notably different premise vis-\`{a}-vis survey-based research, since data is largely given by what a platform stores and what is made available to the researcher -- hence, the theoretical fitting often happens post hoc~\citep{Howison2011}.
A major difficulty is that in order to avoid measurement errors, instead of designing a survey instrument, digital trace studies must consider several steps of the process at once: (i) definition of a theoretical construct aligned with an ideal measurement (ii) the platform to be chosen, (iii) ways to extract (subsets of) data from the platform, (iv) ways to process the data, and (v) the actual measurement of the construct through manifest indicators extracted from the data~\citep{lazer2015issues}, (vi) understand \textit{for whom or what} conclusions are being drawn and how that population relates to the traces observed in the digital traces~\citep{jungherr2017normalizing}.

%Working with digital traces also requires reflecting on representation errors in all process steps. Large differences in regard to representation compared to surveys lie in (i) the ex-ante given sampling frame that is restricted by the platform and usually induces a large coverage error if the target population is off-platform (non-probabilistic), (ii) the lack of a specific request stimulus in found data (non-reactive) and (iii) the described uncertainties regarding traces and users.

%\eg{Motivating Example: Measuring Presidential Approval}{We use the scenario of measuring presidential approval through surveys and digital trace data as a motivating example for both, i) finding the commonalities and dissimilarities of the two paradigms and ii) illustrating our error framework. In a survey [...describe survey of PA]}

\subsection{Ethical Challenges. }%The TED framework, like the TSE, is aimed at identifying and characterizing error sources. 
For a more comprehensive reflection than we can offer here, we point towards the growing body of research specializing on ethical dimensions of working with data from web and social media, such as work by~\citet{zimmer2017internet} and ~\citet{olteanu2016social}. There are also initiatives to provide researchers with practical guidance for ethical research design (e.g.,~\citet{AOIR2019Internet} for the Association of Internet Researchers). In our context it is important to note that applying ethical decision-making might potentially limit options in research design -- which could contribute to error sources in some cases. For example, research is already often based on only those (parts of) platform data which are publicly available, which may add to errors on the platform selection and data collection levels. Users have typically not consented to be part of a specific study and may not be aware that researchers are (potentially) utilizing their data~\cite{fiesler2018participant}, but greater awareness of research activities may also lead to changing user behaviour (or specific groups of users withdrawing from certain activities). Although research in this area is typically not targeted at individual users but rather large collections of aggregated user data, research designs may still have ethical impact on certain groups (or even on the individual level). For example, automatic classification of user groups with machine learning methods can lead to reinforcement of social inequities, e.g. for racial minorities (see below for ``user augmentation error''). In the future it will be interesting to see how new approaches for applying informed consent practices to digital trace data can be adapted, e.g. via data donations. 
It has also been shown that it may not always be possible to prevent de-anonymizing attempts, which might even aggravate as research procedures and their potential errors are becoming more open and being described in greater detail.

\section{A Total Error Framework for Digital Traces of Human Behavior on Online Platforms (TED-On)}

In the remainder, we map the different stages of research using digital traces with those of a survey-based research pipeline where appropriate. Finding equivalencies between the stages facilitates examining errors in digital trace based analysis through the lens of the TSE. Finally, we account for the divergences between the two paradigms and accordingly adapt the error framework to describe and document the different kinds of measurement and representation errors lurking in each step (Figure \ref{fig:data_coll}).

\subsection{Construct Definition}\label{sec:construct_definition}

%\begin{comment}

%\section{An Overarching Concern: Construct Validity}

%The first step in any social science study requires defining the theoretical construct of interest and establishing a theoretical link between the captured data and the construct~\citep{Howison2011}. 
%a theory that links the construct of interest to a valid indicator~\cite{}.
% add some refs from Jungherr paper e.g., Carmines & Zeller, 1979; Cronbach & Meehl, 1955; Kane, 1992; Messick, 1989

%Construct validity refers to whether an indicator (i.e., scale, a test or a metric) is an effective measure of the theoretical construct. Construct validity must be considered in any stage of digital trace based study and we will explain in the later section how different biases may impact construct validity at any stage.

(Theoretical) Constructs are abstract `elements of information'~\citep{groves2011survey} that a scientist attempts to measure by recording responses through the actual (survey) instrument and finally, by analysis of the responses. 
A non-definition or under-definition of the construct or a mismatch between the construct and the envisioned measurement corresponds to issues of \textbf{\emph{validity}}. Given that digital trace data from any particular platform is not specifically produced for the inferential study, researchers must establish a link between the behavior that is observable on the platform and the theoretical construct of interest. 
The first step of transforming a construct into a measurement involves defining the construct; this requires thinking about competing and related constructs, in the best case rooted in theory. Next, researchers have to think about how to operationalize the construct. This entails deliberating about whether a potential ideal measurement sufficiently captures the construct from the given data and if the envisioned measurement does not also – or instead – capture other constructs, i.e., think about convergent vs. discriminant validity of a measurement ~\citep{jungherr2017digital}. 

Since digital trace based studies might proceed in a non-linear fashion, researchers may or may not begin their study with a theoretical construct and a pre-defined measurement in mind. Sometimes a researcher might start with a construct but reevaluate it, and the corresponding measurement, throughout the study depending on the nature of the digital traces. Since the data is largely given by what the platform/system stores, what is available for the public and/or what can be accessed via Application Programming Interfaces (APIs), it may require rethinking the original construct and its definition. Another alternative, which~\citet{salganik2017bit} describes as the ``ready-made'' approach, is to start with a platform or dataset, and then envision constructs that can be studied from that particular platform or dataset.

%\addlinespace
\eg{Example: Construct Definition}{An example of a construct that researchers have been aiming to measure with digital trace data is presidential approval. Whereas in a survey one expresses the defined construct as directly as possible in the questionnaire (``Do you approve or disapprove of the way Donald Trump is handling his job as president?"\footnote{The survey question for presidential approval has remained largely unchanged throughout the years since its inception: https://news.gallup.com/poll/160715/gallup-daily-tracking-questions-methodology.aspx}) a digital trace data researcher may consider \textbf{the act of tweeting about the president positively to be equivalent to approval}~\citep{o2010tweets,pasek2019s}. While positive mentions may indicate approval, researchers should also investigate whether the tweets focus on the presidential role and in fact constitute the ideal -- or at least an appropriate -- measurement. Due to the unsolicited nature of Twitter data, it can be difficult to disentangle if the tweets are targeted towards presidential or personal activities or features. E.g., while comments about the president's private life may indirectly impact approval ratings, they do not directly measure how the president is handling his job, thereby weakening the measurement.}

\subsection{Platform Selection}\label{sec:platform_selection}

In selecting a platform, the researcher needs to ensure the general existence of a link between digital traces that are observable on the platform and the theoretical construct of interest. They, however, also need to account for the impact of the platform and its community on the observable traces and the (likely) divergence between the target population of interest and the platform population. Below, we discuss the errors that may occur due to the chosen platform(s).

\textbf{Platform Affordances Error.} Just as in survey design, content and collection modes may influence responses,\footnote{For example, questions can be posed on topics for which there are more and less socially desirable responses and data collection modes can promote or inhibit accurate responding.} the behavior of users and their traces are impacted by (i) platform-specific sociocultural norms as well as (ii) the platform's design and technical constraints~\citep{wu2020platform}, leading to measurement errors which we together summarize as \textit{platform affordances error}.

For example, Facebook recommends ``people you may know'', thereby impacting the friendship links that Facebook users create~\citep{malik2016identifying}, while Twitter enforcing a 280-character limit on tweets influences the writing style of Twitter users~\citep{gligoric2018constraints}. \enquote{Trending topic} features can shift users' attention, and feedback buttons may deter utterances of polarizing nature. 
Also, perceived or explicated norms -- such as community-created guidelines or terms of service -- can influence what and how users post; for example politically conservative users being less open about their opinion on a platform they regard as unwelcoming of conservative statements, or strict moderation and banning being avoided by contributors through self-censoring.\footnote{Emerging, ideologically extreme platforms like Gab and Parler have the potential to lead to ``migration'' of whole user groups away from mainstream platforms, polarizing the platform landscape.~\citep{ribeiro2020does}}
Similarly, perceived privacy risks can influence user behavior. A major challenge for digital trace-based studies is, therefore, disentangling what~\citeauthor{ruths2014social} call ``platform-driven behavior'' from behavior that would occur independently of the platform design and norms.

Evolving norms or technical settings may also affect the validity of longitudinal studies~\citep{bruns2016twitter} since these changes may cause ``system drifts'' as well as ``behavioral drifts''~\citep{salganik2017bit}, contributing to reduced reliability of measurement over time~\citep{lazer2015issues}. Researchers should thoroughly investigate how particular platform affordances may affect their envisioned measurement.

\vspace*{5px}
\eg{Example: Platform Affordances}{Platform norms such as the character limit, platform design and features (e.g. display of trending topics or posts, feedback buttons) or terms of service and cultural norms can inhibit how and to what extent users express their opinion about the president, leading to platform affordances error. For example, users may have to write terse tweets or a thread consisting of multiple tweets to express their opinion on Twitter. Users may also be less likely to expose an opinion that they predict to be unpopular on the platform, for instance if platforms only allows positive feedback via like-buttons.}

\textbf{Platform Coverage Error.} The mismatch between the individuals in the target population and those being represented on the platform is the \textit{platform coverage error}, a representation error. It is related to coverage error in the TSE, as the given sampling frame of a platform is generally not aligned with the target population (unless the platform itself is being studied). %It is set ex-ante and unchangeable by the researcher for a given platform. 
Further, different online platforms exhibit varying inclusion probabilities, as they attract specific audiences because of topical or technological idiosyncrasies~\citep{smith2018social}.
\footnote{\label{fn:twitterdemo}http://www.pewinternet.org/2018/03/01/social-media-use-in-2018/} 
Twitter's demographics, as a particular example, tend to be different from population demographics~\citep{mislove2011understanding}, and Reddit users are predominantly young, Caucasian and male~\citep{duggan20136}. Population discrepancies could also arise due to differences in internet penetration or social media adoption rates in different socio-demographic or geographical groups, independent from the particular platform~\citep{Wang:2019:DIR:3308558.3313684}. The change of platforms' user composition over time has additionally to be taken into account in terms of reliability of the study~\citep{salganik2017bit}.\footnote{Duplicate users for a member of the target population may also occur depending on the selected platform.}

Lastly, platform affordances also affect coverage: the inhibition of users' traces to the point that they do not produce \textit{any} relevant traces for the research in question effectively makes them ``non-respondents''. This, for all intents and purposes, is indistinguishable from platform coverage error~\citep{eckman2017undercoverage}.

\vspace*{5px}
\eg{Example: Platform Coverage}{ Assume that full access to all relevant data on Twitter is given. 
Then, the sampling frame is restricted to users who have \textit{self-selected} to register and to express their opinion on this specific social media platform. These respondents are likely not a representative sample of the target population (e.g., US voters) which causes a platform coverage error that may lead to highly misleading estimates~\citep{fischer1995desirability}. In addition, 
non-humans (e.g., bots, businesses) act as users and are therefore part of the sampling frame that is defined by the platform.}

A platform coverage error can be thought of as a counterpart to %nonresponse and sampling error, 
coverage error in the TSE. Researchers, as in survey methodology, may reweight participants by socio-demographics (directly available or inferred) to potentially obtain a representative sample~\citep{pasek2018stability,locker1993effects,an2015whom} (see Section~\ref{sec:analysis}), though the efficacy of these correction methods depends on the nature of the self-selection of users~\citep{schnell2017differences}.

\subsection{Data Collection}\label{sec:data_collection}

\FloatBarrier

\begin{table*}[!ht]
\centering
\caption{Different types of feature-based data collection strategies, their explanation and example.}
\label{tab:data_coll}
\resizebox{\textwidth}{!}{
\begin{tabular}{@{}lll@{}}
\toprule
\textbf{Query Type} & \textbf{Definition}                                                                                                                             & \textbf{Examples of Research Question}                                                           \\ \midrule
Keyword  & \begin{tabular}[c]{@{}l@{}}Using keywords including terms, hashtags, image tags\\ regular expressions to subset traces such as posts \\ (tweets, comments, images) or users (users) \end{tabular} & \begin{tabular}[c]{@{}l@{}}Predicting Influenza rates from search queries~\citep{yuan2013monitoring}\\ Understanding the use and effect of psychiatric drugs \\through Twitter~\citep{buntain2015your}\end{tabular} \\
\toprule
Attribute  & \begin{tabular}[c]{@{}l@{}}Using attributes such as location or community\\ affiliation to subset users such as users who may have\\ the relevant attribute in their biography\end{tabular} & \begin{tabular}[c]{@{}l@{}}Inferring demographic information through mobility patterns\\ on photosharing platforms~\citep{riederer2015don}\\Geographic Panels used to study responses to mass shootings\\and TV advertising~\citep{zhang2016geolocated} \end{tabular} \\ 
\toprule
\begin{tabular}[c]{@{}l@{}}Random Digit\\Generation\end{tabular}  & \begin{tabular}[c]{@{}l@{}}Generating random digits and using them as identifiers\\of platform users\end{tabular}                                                    & \begin{tabular}[c]{@{}l@{}}Studying collective privacy behaviour of users~\citep{garcia2018collective}\\ Understanding the demographics and voting behaviour of\\Twitter users~\citep{barbera2016less}\end{tabular}      \\
\toprule
Structure  & \begin{tabular}[c]{@{}l@{}}Using structural properties of users or traces to select\\ data such as interactions (retweeting, liking, friending)\end{tabular} & \begin{tabular}[c]{@{}l@{}}Understanding the influence of users~\citep{cha2010measuring}\\ Predicting political affinity of Twitter users based\\ on their mention networks~\citep{conover2011predicting}\end{tabular}      \\
\toprule
%Lists   & Using lists or subcommunities of users or traces                                                                                                        & \begin{tabular}[c]{@{}l@{}}Using Wikipedia page views to study Influenza \\prevalence~\citep{mciver2014wikipedia}\\ Studying hateful content on Reddit~\citep{chandrasekharan2017you}\end{tabular}        \\ \bottomrule
\end{tabular}}
\end{table*}

After choosing a platform, the next step of a digital trace based study consists of collecting data, e.g., through official APIs, web scraping, or collaborations with platform/data providers. Then, even if the full data of recorded traces is in principle available from the platform, researchers often select a subset of traces, users or both, by querying the data based on explicit features These can, for instance, include keywords or hashtags~\citep{o2010tweets,diaz2016online, stier2018systematically}, and for users, location or demography markers, or their inclusion in lists~\citep{chandrasekharan2017you}, see Table \ref{tab:data_coll}.
This process is usually implemented (i) to discard traces (e.g., tweets) that are presumed to be irrelevant to the construct or (ii) to discard users (e.g., user profiles) that are unrelated to the elements in the target population. 
Additionally, if a researcher selects traces, they may have to collect supplementary information of the user leaving the trace (for example, profile and other posts of the user authoring the tweet)~\citep{schober2016social}. We discuss researcher-controlled selection of traces and users below, and afterwards cover the case of provider-restricted data access.

\textbf{Trace Selection Error. } Typically, researcher-specified queries are used to capture traces broadly relevant to the construct of interest, to be possibly filtered further at later stages. These query choices determine on what basis the construct of interest is quantified and thus may lead to measurement errors. The difference between the ideal measurement and traces collected due to the researcher-specified query is termed \textit{trace selection error}. The trace selection error is loosely related to measurement error\footnote{Not all ``measurement errors'', but the specific ``measurement error'' box on the left side of Fig. \ref{fig:tse}} in the TSE -- the difference between the ideal, truthful answer to a question and the actual response obtained in a survey -- in that traces might be included which do not truly carry information about the construct, or that informative ones might be missed. 
Low precision and low recall of queries may directly impact the measurement which can be established from the subselection of platform data used. He and Rothschild examine different methods of obtaining relevant political tweets, establishing that bias exists in keyword-based data collection~\citep{he2016selection} affecting both the users included in the sample as well as the sentiment of the resultant tweets.

\vspace*{5px}
\eg{Example: Trace Selection}{ Assume that we aim to capture all queries about the US president entered into a search engine by its users. If searches mentioning the keyword ``Trump'' are collected, searches which refer to Melania Trump or ``playing a trump card'', could be included and lead to noise. Likewise, relevant queries might be excluded that simply refer to ``the president", ``Donald" or acronyms, which might in turn be linked to specific user groups (e.g., younger people). (Not) excluding (in)eligible traces in this way might severely harm the soundness of the measurement.} %(which we discuss in more detail in Section ``Analysis and Inference''). 
\vspace*{5px}
One may assess signal selection error through analysis of precision and recall of the queries being used~\citep{ruiz2014efficient}, and mitigate them via query expansion \citep{linder2017improved}. While low precision can usually be addressed in subsequent filtering steps after the data collection is finished (see Section ``Data Preprocessing''), the non-observation of relevant signals cannot be remedied without repeating the data selection step.

\textbf{User Selection Error. } While traces are selected according to their estimated ability to measure the construct of interest, their exclusion usually also removes those users that produced them from the dataset (think of Twitter profiles attached to tweets), if no other traces of these users remain in the subset.\footnote{This is typically true for most data collection strategies except Random Digit Generation where users are selected independently of their characteristics.} In this manner, users with specific attributes might be \textit{indirectly} excluded simply because they produce traces that are filtered out (e.g., teenagers may be excluded because they use different terms to refer to the president).
The error incurred due to the gap between the selection of users and the sampling frame is called the \textit{user selection error}. It is a representation error related, but distinct from, the sampling error in the TSE. It is also related to the coverage error if one considers the researcher-specified query result set as a second sampling frame,\footnote{A collection of data through explicit features is equivalent to defining a boundary around the traces or users to be used for analysis further on~\citep{gonzalez2014assessing}.} where user selection error is the gap between the sampling frame enforced by the query and the platform population.
This error \textit{also} occurs if users are \textit{directly} excluded by their features (not via their traces); for instance, by removing user-profiles deemed irrelevant for inferences to the target population as determined by their indicated location, age or (non-)placement on a user list. This is especially critical since the voluntary self-reporting of such attributes differs between certain groups of %individuals that are behind the collected 
users, such as certain demographics being less prone to reveal their location or gender in their account information or posts~\citep{pavalanathan2015confounds}, and additionally can be unreliable due to their variation over time \citep{salganik2017bit}.

There are many \textbf{approaches to collecting data} for both traces and users (cf. Table \ref{tab:data_coll}) and each comes with different types of trace and user selection pitfalls. As keyword-based search is a popular choice, Tufekci analyzed how hashtags %\footnote{hashtags can be thought of as crowdsourced keywords or tags~\citep{olteanu2016social}}
can be used for data collection on Twitter and finds that hashtag usage tapers down as time goes on -- users continue to discuss a certain topic, they merely stop using pertinent hashtags~\citep{tufekci2014big}. For collecting data related to elections or political opinion, many studies use mentions of the political candidates to collect data related to them~\citep{barbera2016less,o2010tweets,diaz2016online, stier2018systematically}. While this a high-precision query, it may have low recall, excluding users who refer to political candidates with nicknames, thereby reducing the sample's generalizability. On the other hand, it may also include ineligible users who might have been referring to someone with the same name as the politician, in case a candidate's name is common.

In addition to keyword selection, other sources for data selection are also in use, e.g., those based on attributes, such as location~\citep{bruns2014twitter} or structural characteristics~\citep{demartini2007finding} or affiliation to particular subcommunities or lists~\citep{chandrasekharan2017you}, and random digit generation (RDG)~\citep{barbera2016less}.
%\footnote{Network sampling strategies are usually non-probabilistic.} 
Each method has different strengths and weaknesses, with random digit generation (RDG) being closest to the Random Digit Dialing method of conducting surveys, although the method may generate a very small sample of relevant tweets~\citep{barbera2016less}.\footnote{Recent changes to Twitter's ID generation mechanism renders the RDG data collection method no longer viable: \url{https://developer.twitter.com/en/docs/basics/twitter-ids}} Lists of users as well as selection based on reported attributes also restrict the dataset to users who have either chosen to be marked by them, have been marked by others, often leading to additional 
coverage error since the characteristics of selected users may be systematically different from the target population~\citep{cohen2013classifying}.
Further, when network data is collected, different crawling and sampling strategies may impact the accuracy of various global and local statistical network measures such as centrality or degree~\citep{Galaskiewicz1991,Borgatti2006,Kossinets2006,wang2012measurement,Lee2015,costenbader2003stability}, the representation of minorities and majorities in the sample network~\citep{wagner2017sampling} and the estimation of dynamic processes such as peer effects in networks~\citep{yang2017should}. This is especially important if the construct of interest is operationalized with structural measurements (e.g., if political leaning is assessed based on the connectivity between users and politicians or if extroversion is assessed based on the number of interaction partners).

Additionally, \textbf{data access restrictions by platform providers} can aggravate trace and user selection errors. 
Many platforms regulate data access to a provider-determined subset that may not be a probabilistic sample of the whole set of traces in their system -- inducing a representation error, but in this instance regarding the \textit{platform population}. Such an artificial restriction can indeed be closely linked to \textit{sampling error} in the TSE, since a nonprobabilistic sample is drawn from the platform, though not by the researcher. 
For example, while for some platforms like Github or Stack Exchange, the entire history of traces ever generated are available and the researcher can make selections from this full set freely, others like Facebook Ads or Google Trends only share aggregated traces. 
As the most prominent example, Twitter provides data consumers with varying level of access. A popular choice of obtaining Twitter data is through the 1\% `spritzer' API, yet research has found that the free 1\% sample is significantly different from the commercial 10\% `gardenhose' API~\citep{morstatter2013sample}. 
One reason for a misrepresentation of users is overly active users being assigned a higher inclusion probability than others in such a sample, simply because they produce more traces and sampling is done over traces in a given time frame~\citep{schober2016social}.

 Both platforms and researchers may also decide to limit what data can be collected based in order to protect user privacy.\footnote{Platforms deal with content deletion by users in different ways, which can also affect the resultant subsets, a point we hope to explore in future work.} 
%Irrespective of whether full plat is available or not, researchers further sample based on certain features of traces or users that is pertinent to their research question. Researchers may also sample users from the platform census due to high computational costs of preprocessing the entire census. 

\vspace*{5px}
\eg{Example: User Selection}{When studying political opinions on Twitter, vocal or opinionated individuals' opinions will be overrepresented, especially when data is collected based on traces (e.g., tweets), instead of individual accounts (e.g., tweets stratified by account activity)~\citep{barbera2016less,o2010tweets,pasek2018stability,diaz2016online} simply because they tweet more about the topic and have a higher probability of being included in the sample than others.
Further, certain groups of users (e.g., teenagers or Spanish speaking people living in the US) may be underrepresented if keyword lists are generated that mainly capture how adult Americans talk about politics.}

\subsection{Data Preprocessing}\label{sec:data_preprocessing}
Data preprocessing refers to the process of removing noise in the form of ineligible traces and users from the raw dataset as well as augmenting it with extraneous information or additionally needed meta-data. 

\subsubsection{Trace Preprocessing}
Trace preprocessing is done since researchers may want additional information about traces or may want to discard ineligible traces that have been mistakenly included due to data collection. The reduction or augmentation of traces, while aimed at improving our ability to measure the construct through them, may also further distort them. 

\textbf{Trace Augmentation Error.}
%such as textual posts or ``likes'', 
The augmentation of traces can comprise several means: sentiment detection, named entity recognition, or stance detection on the text content of a post; the annotation of ``like'' actions with receiver and sender information; or the annotation of image material with objects recognized within.
This augmentation is done as part of measuring the theoretical construct and, due to the size and nature of the data, mainly pertains to the machine-reliant coding of content. %, although manual labeling (e.g., of training sets) also falls under this term.
An error may be introduced in this step due to false positives and negatives of the annotation method.
Just as answers to open-ended questions in a survey are coded to ascertain categories and inadequately trained coders can assign incorrect labels, \textit{trace augmentation error} occurs often due to inaccurate categorization due to the usage of pretrained ML methods or other pre-defined heuristics, frequently designed for different contexts. Particularly the annotation of natural language texts has been popular in digital trace studies and even though a large body of research has emerged on natural language processing methods, many challenges remain~\citep{puschmann2018turning}, including a lack of algorithmic interpretability for complex or black-box models. Study designers need to carefully assess the context for which an annotation method was built before applying it and consult benchmarking studies for commonly used methods \citep{gonzalez2015signals}. In case methods have been developed for a different type of content, domain adaptation may also be used to improve the performance of methods trained on a data source different from the current data of interest~\citep{hamilton2016inducing}.

\textbf{Trace Reduction Error.}
Finally, certain traces may be ineligible items for a variety of reasons such as spam, hashtag hijacking or because they are irrelevant to the task at hand. The error incurred due to the removal of ineligible traces is termed the \textbf{\emph{trace reduction error}}. Researchers should investigate precision and recall of methods that remove irrelevant traces to estimate the trace reduction error~\citep{kumar2018false}. %\haiko{This is relatively short.} 

\vspace*{5px}
\eg{Example: Trace Preprocessing}{
\textbf{Augmentation:} 
Political tweets are often annotated with sentiment to understand public opinion~\citep{o2010tweets,barbera2016less}. However, the users of different social media platforms might use different vocabularies (e.g., "Web-born" slang and abbreviations) than those covered in popular sentiment lexicons and even use words in different contexts, leading to misidentification or under-coverage of sentiment for a certain platform or subcommunity on that platform~\citep{hamilton2016inducing}. 
%Typically, most commonly available tools cater to content produced in English. This could be potentially problematic for understanding presidential approval in the USA where the Spanish speaking community may have views that diverge from the English-speaking community. Leaving out content produced in Spanish could lead to biased estimates.

\textbf{Reduction:} Researchers might decide to use a classifier that removes spam, but has a high false positive rate, inadvertently removing non-spam tweets. They might likewise discard tweets without textual content, thereby ignoring relevant statements made through hyperlinks or embedded pictures/videos. 
}

\subsubsection{User Preprocessing}
In analogy to reducing or augmenting traces, users may also be preprocessed. In this step, the digital representations of humans, such as profiles, are augmented with auxiliary information and/or certain users are removed because they were ineligible units mistakenly included in the sampling frame. These preprocessing steps can affect the representativeness of the final data used for analysis.

\textbf{User Augmentation Error.} Researchers might be interested in additional attributes of the users in the data sample, which may serve as dependent on independent variables in their analysis; for example, a research might be interested in how political leaning affects the propensity to spread misinformation. Additional user attributes are generally also of interest for reweighting digital trace data by socio-demographic attributes and/or by activity levels of individuals. The former is traditionally also done in surveys to mitigate representation errors (discussed as `Adjustment' in Section ``Analysis and Inference''). 

However, since such attributes are rarely pre-collected and/or available in platform data, demographic attribute inference for accounts of individuals is a popular way to achieve such a task, often with the help of machine learning methods~\citep{zhang2016your,rao2010classifying,sap2014developing, Wang:2019:DIR:3308558.3313684}. %and removing these users from the dataset is common.
Of course, demographic attribute inference itself is a task that may be affected by various errors ~\citep{Karimi2016,mccormick2017using} and can be especially problematic if there are different error rates for different groups of people~\citep{buolamwini2018gender}. Furthermore, these automated methods are often underspecified, operating on a limited understanding of social categories such as gender~\citep{scheuerman2019computers} which often excludes transgender and gender non-binary people~\citep{hamidi2018gender}. There are other crucial departures from survey research in gender attribution---- automated methods (mis)gender people at scale and \textit{without} their consent~\citep{keyes2018misgendering}, whereas demographic attributes are self-reported in surveys. Similar tensions apply to inferring race~\citep{khan2021one}.
Such automated approaches for classifying users according to (perceived) characteristics thus have a strong ethical component, as they might harm marginalized communities by more frequently misclassifying minorities~\citep{buolamwini2018gender})

Platforms may also offer aggregate information about their user base, which can potentially be supplied through provider-internal inference methods and be prone to the same kind of errors without the researcher knowing~\citep{zagheni2017leveraging}. 
Besides demographic features calculated network position of the account on the platform, or linking to corresponding accounts on other platforms also constitute user augmentation. 

The overall error incurred due to the efficacy of user augmentation methods is denoted as \textbf{\emph{user augmentation error}}. 

%User augmentation error may be quantified based on an error analysis of the methods used for annotating auxiliary information of the data, such as the accuracy of demographic inference. The reliability of multiple augmentation methods can also be assessed~\citep{gonzalez2015signals}.

\textbf{User Reduction Error.} In addition to user augmentation, preprocessing steps usually followed in the literature include removing inactive users, spam users, and non-human users -- comparable to ``ineligible units'' in survey terminology -- or filter content based on various observed or inferred criteria (e.g., location of a tweet, the topic of a message). Similar to user augmentation error, the methods for detecting and removing specific type of users are usually not perfect and the performance may depend on the characteristics of the data (cf. \citep{de2018lobo} and \citep{wu2018twitter}). 
Therefore it is important that researchers analyze the accuracy of the selected reduction method to estimate the \textbf{\emph{user reduction error}}. 
%This error can be the result of researchers not removing ineligible units that contribute to noisy traces or filtering criteria that are too strict. 
%As an approximate counterpart to this error one can consider \upd{coverage} error in surveys. 
%Previous research that compared various methods for bot~\citep{de2018lobo} and spam detection~\citep{wu2018twitter}, which are often used in preprocessing digital trace data, found that different methods have varying performance depending on the characteristics of the data. 

\vspace*{5px}
\eg{Example: User Preprocessing}{
\textbf{Augmentation:}
Twitter users’ gender, ethnicity, or location may be inferred to understand how presidential approval differs across demographic groups. Yet, automated gender inference methods based on images have higher error rates for African Americans than White Americans~\citep{buolamwini2018gender}; therefore, gender inferred through such means may inaccurately estimate approval rates among African American males versus females compared to their White counterparts, which raises serious ethical concerns. 

\textbf{Reduction:}
While estimating presidential approval from tweets, researchers are usually not interested in posts that are created by bots or organizations. In such cases, one can detect such accounts~\citep{alzahrani2018finding,mccorriston2015organizations} or detect first-person accounts as done by~\citet{an2015whom} and remove users who did not posts first-had accounts, even if the first selection process for data collection has retained them.}

%. Another source of user augmentation error could be due to the biases in self-reporting location on Twitter~\citep{pavalanathan2015confounds}.
%User reduction error can be assessed by understanding the criteria or definition chosen for exclusion, and researchers should note if their criteria could potentially exclude users that act as effective ``sensors''. 

%To describe and mitigate preprocessing errors, researchers should analyze if the chosen augmentation and reduction methods are suitable for the particular type of digital traces since different errors may occur for different types of content~\citep{gonzalez2015signals}. In case methods have been developed for a different type of content, domain adaptation may also be used to improve the performance of methods trained on a data source different from the current data of interest~\citep{yang2017overcoming,hamilton2016inducing}.
%Researchers should also pay attention to how different groups of users interact with the platform. For example, systemic biases could be introduced if certain groups of users are more willing to share information that makes metadata inference easier, impacting user augmentation. Additionally, different styles and behaviors may affect how traces are produced (what~\cite{olteanu2016social} describe as \textit{content biases}), which in turn affects trace preprocessing errors since common tools for trace augmentation %(especially sentiment analysis methods) and reduction may have differential errors for different types of content~\citep{gonzalez2015traces}.

\subsection{Analysis and Inference}\label{sec:analysis}

After preprocessing the dataset, an estimate is calculated for the construct in the target population.

%Finally, after having preprocessed the dataset, we move on to measuring a final indicator for the construct of interest. As noted before, depending on the nature and availability of the digital traces, the construct and its corresponding measurement may be defined or redefined at this stage. Note that in this step, the traces are annotated with respect to the construct (say, presidential approval) and is therefore a separate step to trace augmentation, where traces are annotated with extraneous information that may help measure the construct (such as parts-of-speech tags or sentiment). We discuss the errors resulting from this step below.

\textbf{Trace Measurement Error. } 
%After we have obtained the augmented traces, either through manual coding or an automated process, they are aggregated to arrive at the final estimate. Just as the set of users is adjusted through reweighting to correct for representation errors, 

In this step, a concrete measurement of the construct is calculated -- say, a [0,1] value of presidential approval from a vector of sentiment scores associated with a tweet -- and models are built that link this measurement to other variables. It is therefore distinct to trace augmentation, where traces are annotated with \textit{auxiliary} information to measure the construct (parts-of-speech tags, sentiment, etc.). In practice, these steps can be highly intertwined, but it is still useful to conceptually disentangle them as separate stages. 

The researcher can calculate the final estimate through many different techniques, from simple counting and heuristics to complex ML models. While simpler methods may be less powerful, they are often more interpretable and less computationally intensive. %, regarding data and computation needed. 
On top of errors incurred in previous stages (e.g., not all dimensions of the construct captured during data collection), the model used might only detect certain aspects of the construct or on spurious artifacts of the data being analyzed. For example, a ML model spotting sexist attitudes may work well for detecting %gender-specific stereotypes but not well for detecting the rejection of feminism which are both dimensions of the construct sexism 
obviously hostile attacks based on gender identity but fail to capture benevolent sexism, though they are both dimensions of the construct of sexism \citep{jha2017does,samory_call_2021}.
Even if the technique is developed or trained for the specific data collected, it can therefore still suffer from low validity. Any error that distorts how the construct is estimated due to modeling choices is denoted as \textit{trace measurement error}. Even if validity is high for the given study, a replication of the measurement model on a different dataset -- a future collection of the same platform data or a completely different dataset -- can fail, i.e., have low reliability~\citep{conrad2019social, sen2020reliability}. 
Different kinds of traces can be taken into account to different degrees in the final estimate, e.g., in order to account for differences in activity of the users generating the traces (e.g. power users' posts) or their relevance to the construct to be measured. This can lead to erroneous final estimates, which we denote as \textbf{\textit{trace measurement error}}. This error may arise due to the choice of modeling or aggregation methods used by the researcher as well as how the traces are mapped to the users.
%units of observation are mapped to units of analysis.

%The researcher may also face the choice of collapsing all expressions produced by one user into a single trace. If she does not do so, vocal users are given a stronger voice leading to ecological fallacies. On the other hand, errors may also occur based on how traces generated by a single user are aggregated. 

Researchers should also account for heterogeneity since digital traces are generated by various subgroups of users who may have different behavior. Aggregation may mask or even reverse underlying trends of digital traces due to trace measurement errors in the form of Simpson's paradoxes (a type of ecological fallacy~\citep{alipourfard2018can, Howison2011, lerman2018computational}). Further, the performance of machine learning methods will be impacted by this heterogeneity and the performance of a model can differ enormously across different subgroups of users especially if some groups are much larger and/or much more active (i.e. produce more traces that can be used for training).

Trace measurement error has been addressed in related work as well, under different names and to different degrees. For example,
\citet{amaya2020total} as well as \citet{westAnalyticErrorImportant2017} %\mr{as well as (West, Sakshaug.... =$>$as well as West and colleagues (West, Sakshaug...} 
introduce another error, called \emph{analytic error}, which \citet{westAnalyticErrorImportant2017} characterize as \enquote{important but understudied aspect of the total survey error (TSE) paradigm}. Following \citet{amaya2020total}, analytic error is broadly related to all errors made by researchers when analyzing the data and interpreting the results, especially when applying automated analytic methods. A consequence of analytic error can be false conclusions and incorrect models, which manifests in multiple ways (e.g., spurious correlations, endogeneity, causation versus causality) \citep{bakerBigDataSurvey2017}. In that sense, the equivalent of one component of analytic errors, i.e., errors incurred due to the choice of modeling techniques, is the trace measurement error. The other component, errors due to choice of survey weights, is described below in adjustment error and is related to analytic error in the sense of \citet{westAnalyticErrorImportant2017}; in addition, they put special emphasis on applying appropriate estimation methods when analyzing complex samples -- e.g., using proper variance estimation methods.
\vspace*{5px}
\eg{Example: Trace Measurement}{
Depending on the way the construct was defined (say, positive sentiment towards the president), and trace augmentation performance (lexicons which count words with a sentiment polarity), the researcher obtains tweets whose positive and negative words have been counted. Now the researcher may define a final link function which combines all traces into a single aggregate for the construct of ``approval''. She may choose to count the normalized positive words in a day~\citep{barbera2016less}, the ratio of positive and negative words per tweet or add all the ratio of all tweets in a day~\citep{pasek2019s}. %Essentially, the researcher has a choice of methods that she may use, and each method has its own weaknesses.
The former calculation of counting positive words in a day may underestimate negative sentiments of a particular day, while in the latter aggregate, negative and positive stances in tweets which report both would neutralize each other, resulting in the tweet having no sentiment.

A user may have expressed varying sentiments in multiple tweets. The researcher faces the choice of averaging the sentiment across all tweets, taking the most frequently expressed sentiment or not aggregating them at all and (implicitly) assuming each tweet to be a trace of an individual user~\citep{tumasjan2010predicting, o2010tweets, barbera2016less} which may amplify some users' voices at the cost of others.

Either to avoid the vocabulary mismatch between pre-defined sentiment lexica and social media language or because sentiment is an inadequate proxy for approval, researchers may alternatively want to use machine learning methods that learn which words indicate approval towards the president by looking at extreme cases (e.g., tweets about the president from his supporters and critics). The validity of this measurement depends on the data it is trained on and to what extent this data is representative of the population and the construct. That means the researchers have to show that the selected supporters and critics expose approval or disapproval towards the president in a similar way as random users~\citep{cohen2013classifying}.
}

\textbf{Adjustment Error.} To infer to the target population, researchers can attempt to account for platform coverage error (possibly aggravated by user selection or preprocessing). Thus, they may use techniques leveraged to reduce bias when estimating population parameters employing nonprobabilistic samples, such as opt-in web surveys~\citep{goel2015non, goel2017online}. Researchers have suggested specific ways to handle the nonprobabilistic nature of these surveys, which may also be applied to digital traces~\citep{kohler_possible_2019, kohler2019nonprobability}. In contrast to \citet{kohler_possible_2019} or \citet{kohler2019nonprobability}, some researchers suggest weighting techniques like raking or post-stratification to correct for coverage and sampling errors. The availability of demographic information -- or other features applicable for adjustment -- as well as the choice of method can both cause errors which we label \textit{adjustment error}, in line with \citet{groves2011survey}. It should be noted that corrections solely based on socio-demographic attributes are not a panacea to solve coverage or non-response issues \citep{schnell2017differences}. 

%Recently, inspired by \hl{weighting approaches in the} social science 
%\mr{Social science? Do you mean developments in the social sciences of weighting methods? I believe all development of weighting takes place in the social sciences (sometimes in statistics departments, but driven by the particular problems introduced to sampling when people are involved) so this jumped out at me as an odd choice of words. Please elaborate and clarify.}
Reweighting has been explored in studies using digital traces~\citep{zagheni2015demographic,barbera2016less,pasek2018stability,pasek2019s, Wang:2019:DIR:3308558.3313684} using calibration or post-stratification. Unlike in survey methodology where scholarship has systematically studied which approaches work well for which type of data sources~\citep{cornesse2020review,kohler_possible_2019, kohler2019nonprobability,cornesse2020response}, there is a lack of comprehensive study regarding reweighting in digital traces. Broadly, there are two approaches to reweighting in digital trace based studies. The first approach reweights the data sample according to a known population, for example obtained through the census distribution~\citep{yildiz2017using,zagheni2015demographic}.
The second approach reweights general population surveys according to the demographic distribution of the online platform itself ~\citep{pasek2018stability,pasek2019s} -- found through, e.g., usage surveys or provided by the platform itself. While the second method has the advantage of bypassing the use of biased methods for demographic inference (thus mitigating user augmentation errors), the platform demographics might not apply to the particular dataset since all users are not equally active on all topics and the error introduced by this reweighting step is difficult to quantify. For the first method, while researchers often use biased methods for demographic inference, the errors of these methods can be quantified through error analysis.

\vspace*{5px}
\eg{Example: Adjustment}{When comparing presidential approval on Twitter with survey data,~\citet{pasek2019s} reweight the survey estimates with Twitter usage demographics but fail to find alignment between the two measures.\footnote{Cf. footnote \ref{fn:twitterdemo} in Section ``Platform Selection''~\ref{sec:platform_selection} for Twitter's demographic composition.} In this case, the researchers assume that the demographics of Twitter users is the same as for a subset of Twitter users tweeting about the president, an assumption which might not be true. Previous research has shown that users who talk about politics on Twitter tend to have different characteristics than random Twitter users~\citep{cohen2013classifying} and that they tend to be younger and have a higher chance of being white men~\citep{bekafigo2013tweets}. Therefore using Twitter demographics as a proxy for politically vocal Twitter users may lead to adjustment errors.}

\section{Application of the Error Framework}\label{sec:case_studies}

 To illustrate the applicability and the utility of the framework, we will now look at typical computational social science studies through the lens of our error framework. 
%Future work will include further case studies to showcase the applicability of our framework. 
%We selected cases that use digital trace data collected from two different sources (Wikipedia and Facebook) to show that the framework is not restricted to specific types of data.  

%
%In the following we discus in computational social science, where computational methods have been applied to studying sociological phenomena. 
We note that the point of this section is not criticize the design choices of the researchers conducting these respective studies, but rather to systematically highlight the errors that can lurk in various steps, most of which the authors have thought about themselves and have applied mitigation approaches to. No study is without limitations, and some errors simply cannot be fully mitigated with current techniques (such as imperfect demographic inference). 
In that vein, a central aim of our error framework is  to systematically describe and document the errors that may occur when designing and conducting digital trace based studies using a vocabulary that enables the communication across disciplines. 

\begin{comment}
\notes{
Running example:\\
Construct: Presidential approval\\
Target Population: American people\\
Frame Population: Twitter users\\
Unit of Observation / Traces: Tweets with a certain attribute\\
Unit of Analysis / Users: Twitter users\\
%Traces: Tweets with a certain attribute\\
%Users: Individual Twitter users\\
}
\end{comment}

\subsection{Case Study: Assessing County Health via Twitter}

The first case study we explore is Culotta's endeavour to infer multiple county-level health statistics (e.g.,
obesity, diabetes, access to healthy foods) based on tweets originating in those counties~\citep{culotta2014reducing}. The author collects tweets which include platform-generated geocodes of 100 American counties, analyzes the content of the tweets using lexicons to infer health-related traces, and finally aggregates them by county. The author also annotates the demographics of the Twitter users and reweights the estimates according to the county statistics to reduce the representation bias of Twitter. In the following section, we outline the different errors possibly occurring in each stage of this study.

\begin{comment}

\notes{
Construct: Health status\\
Target Population: American people\\
Sampling Frame: Twitter users\\
Unit of Observation / Traces: Geocoded tweets\\
Users: Twitter users\\
Unit of Analysis (For whom conclusions are drawn): Counties\\
%Traces: Tweets with geocodes\\
}
\end{comment}

\smallskip 

\subsubsection{Construct Definition and Target Population}
  \hfill\\

\noindent
\textbf{Constructs}: 27 county-level health statistics (e.g.,
obesity, diabetes, access to healthy foods, teen birth rates) are the constructs to be quantified and the ideal measurement envisioned is tweets related to these health issues.
\smallskip

\noindent
\textbf{Target population}: the population of the 100 most populous counties in the USA.

\smallskip 

\noindent
 \textbf{Issues of Validity}: it is unclear if and how people reveal health-related issues on Twitter. Some health-related issues may be more sensitive than others, therefore Twitter users may be less likely to publicly disclose them. Further, it is possible that only certain type of Twitter users reveal health-related issues.
 Lastly, the act of mentioning an issue is not clearly conceptually linked to either experiencing, observing or simply thinking about it. 
  
 \smallskip 
 \smallskip 
 
  \subsubsection{Platform Selection}
  \hfill\\
  
\noindent
   \textbf{Sampling frame:} Twitter accounts and their tweets
   \smallskip 
   
\noindent
  \textbf{Platform Affordances Error:}  Twitter's formatting guidelines (140 character limit at the time of the study), allowances (use of hashtags, links, or mentions) and content sharing restrictions~\footnote{https://help.twitter.com/en/rules-and-policies/media-policy} directly impact how users tweet about health and might limit the expressivity of posts talking about the constructs at hand.
   \smallskip 

\noindent
  \textbf{Platform Coverage Error:} Twitter's skewed demographics present a recurrent problem prone to causing platform coverage error, since the platform population deviates from the target population. The author strives to adjust for this coverage mismatch through reweighting which we discuss below.

    \smallskip 
    \smallskip 
\subsubsection{Data Collection}
  \hfill\\
  
\noindent
The author chooses an attribute-based and trace-based data collection strategy where the attribute of interest is location. Specifically, all tweets corresponding to US county geolocations are collected.\footnote{This design is no longer possible as Twitter has discontinued the use of geolocations: \url{https://twitter.com/TwitterSupport/status/1141039841993355264}} The profile information of the authors of the geotagged tweets is also collected.
   \smallskip 
   
\noindent
\textbf{Traces:} Tweets
     \smallskip 
     
\noindent
\textbf{Trace Selection Error:} Using geolocations to select tweets leads to certain issues on the trace level: (1) Tweets from within the county about the health issues of interest might not be geo-coded and therefore not selected, and (2) some tweets might not reflect conditions in the specific county since they are emitted by users not residing there. 
%users may be physically in the county while tweeting but do not live there and (2) users that live in the county may tweet about health-issues without geo-coding the tweet.
%that are related with individuals that live in a certain county and are therefore part of the health-statistic of this county 

     \smallskip 
     
\noindent
\textbf{Users:} Twitter accounts
     \smallskip
     
\noindent
\textbf{User Selection Error:} Firstly, the above-mentioned trace selection error of course is directly reflected in the erroneous inclusion of  accounts of individuals who are physically in the county at the time of tweeting but do not reside there permanently. Secondly, not all residents who are twitter users use geocodes; in fact, only a part of all users does, which are in turn not representative of the Twitter user population~\citep{pavalanathan2015confounds}.

    \smallskip 
        \smallskip 
\subsubsection{Data Preprocessing}
  \hfill\\
  
\noindent
To quantify health-related traces in the tweets, the author uses two lexicons --  LIWC and PERMA -- to annotate various linguistic, psychological, social and emotional aspects of the tweets as well as the profile descriptions of the tweet authors. Preprocessing steps also include the use of automatic demographic inference for augmenting gender and ethnicity (race) information of the authors of the health-related tweets.
 %\notes{How the target population is defined, determines how adjustment is done and its efficacy in reducing representation errors.}
 
      \smallskip 
      
\noindent
\textbf{Trace Augmentation Error:} Occurs due to shortcomings of lexicon-based approaches. Lexicons are often not designed for social media terminology (such as hashtags) and may therefore suffer from low precision and recall when annotating a tweet with a concept/construct or auxiliary features to measure the construct. 
%the mismatch between words used in a tweet and the underlying emotion of the author~\citep{beasley2015emotional}.
        \smallskip 
        
\noindent
\textbf{User Augmentation Error:} The methods used for demographic inference are likely to introduce at least some inaccuracies, particularly if they are based on a gold standard dataset that does not align completely with the characteristics of the annotated data. 
      \smallskip 
      
\noindent
\textbf{User Reduction Error:} Users for whom demographics cannot be inferred are not included in the analysis. These accounts might systematically be from a certain demographic stratum, e.g., females or older individuals.

%\notes{Adjustment is the process of weighting users based on different attributes including activity.} 
    \smallskip 
        \smallskip 
\subsubsection{Data Analysis}
  \hfill\\
  
\noindent
Finally, the author aggregates the lexica-annotated tweets and user descriptions by county. He also applies reweighting to address the coverage mismatch of Twitter users and the target population.
      \smallskip 
      
\noindent
\textbf{Trace Measurement Error: }The author notes that if certain users are more active than others, their health traces will be weighted accordingly. Therefore, he normalizes by users, mitigating trace measurement error.
      \smallskip 
      
\noindent
\textbf{Adjustment errors:} The choice of method as well as the variables used for reweighting may lead to adjustment errors. For example, gender and ethnicity may not be sufficient attributes to explain the self-selection bias of Twitter users. Furthermore, the coverage bias induced by selecting only those users who have enabled geotagging, is not addressed. 
%The coverage bias induced due to only using geotagged tweets

\smallskip
\smallskip

\subsection{Case Study: Nowcasting Flu Activity through Wikipedia Views}

Our second case study focuses on predicting the prevalence of Influenza-like Illnesses (ILI) from Wikipedia usage by ~\citet{mciver2014wikipedia}, which stands as an illustrative example for a line of research on predictions based on  digital trace data that is based on aggregated traces. Traditionally, cases of ILI  in the US would be measured by reports from local medical professionals, and then be recorded by a central agency like the Center for Disease Control and Prevention (CDC). The authors investigate whether Wikipedia usage rates would be an adequate replacement for reports by the CDC. Specifically, the authors use views on selected Wikipedia articles relevant to ILI topics to predict ILI prevalence.

\smallskip 
\subsubsection{Construct Definition and Target Population}
  \hfill\\
  
\noindent
\textbf{Construct:}  A case of ILI of a person as diagnosed by the CDC.

\smallskip
\noindent \textbf{Target Population:} Described to be the "American Population", which could refer to all U.S. citizens residing in the U.S., all U.S. residents or simply the same population for which CDC data is collected that is used as the gold standard set.
\smallskip

\noindent \textbf{Issues of Validity:} What does the act of accessing ILI-related Wikipedia pages imply about the person who looks at these pages? Do we assume that these persons suffer from influenza or related symptoms themselves, or could this also be individuals with a general interest in learning about the disease, potentially also inspired by media coverage of a related topic? Given that we do assume the majority of readers of  ILI-related articles on Wikipedia to be infected or at least affected individuals: feeling sick does not necessarily imply that a viewer contracted an ILI and not another illness. Only a medical expert can discern ILI-symptoms  from similar diseases.
To test  validity in such cases, one option would be to survey a random sample of Wikipedia readers that land on the selected ILI-related articles to find out about their health status or alternative motivations. Further surveys and/or focus groups could help to learn about if, how, and where people search for advice online when being sick.

\smallskip
\smallskip

\subsubsection{Platform Selection}
  \hfill\\
  
\noindent
\textbf{Sampling frame: } Wikipedia pageviews. The researcher has no access to  accounts linked to  pageviews and can therefore only select a data subset on the trace level. 

\noindent
\textbf{Platform Affordances Error:} The way users arrive at and interact with the articles presented, thereby affecting viewing behavior, is shaped by Wikipedia's interface layout, the inter-article link structure and external search results linking into this structure. I.e., users are frequently arriving at Wikipedia articles from a Google search~\citep{mcmahon2017substantial, dimitrov2019different} implying that Google's ranking of Wikipedia`s ILI-related articles with respect to certain search queries impacts the digital traces we observe in Wikipedia data as well. Past research has also found that article structure and the position of links play an important role in the navigation habits of users~\citep{lamprecht2017structure}.
\smallskip

\noindent
\textbf{Platform Coverage Error:} The readers of Wikipedia are not representative of the target population.

\smallskip
\smallskip

\subsubsection{Data Collection}
\hfill\\

\noindent
The authors of the paper curate a list of 32 Wikipedia articles that they identified to be relevant for ILI, including Avian influenza, Influenza Virus B, Centers for Disease Control and Prevention, Common Cold, Vaccine, Influenza.

\noindent
\textbf{Traces:} Hourly aggregated article views of 32 selected articles, obtained from http://stats.grok.se (which is no longer operable).

\noindent
\textbf{Trace Selection Error:} The views for certain selected articles might not indicate a strong primary interest in ILI (`Centers for Disease Control and Prevention', `Vaccine' are viewed for many other reasons). Other relevant articles' views are not counted -- `influenza vaccine', for example, seems relevant but is not included. 
\smallskip

\noindent
\textbf{Users:} Not observable. Could be assumed to be single browser agents connecting to the server and visiting pages. Each session is an anonymous or a logged in user; however, these are not made available publicly by Wikimedia. 

\noindent
\textbf{User Selection Error: }Assuming that the selected articles might in fact be primarily viewed out of interest in ILI, there remains another potential error source: Different subgroups of the target population (e.g., medically trained individuals) might turn to different articles for their information needs about ILI that are not included in the 32 selected pages (e.g. articles with technical titles).  As these traces are excluded, so are these members of the target population.

A detailed explanation for the selection criteria of the particular set of Wikipedia articles would be useful for comparing the results with related approaches. For example, it could be mentioned if the selection process was informed by specific theory or by empirical findings from interviews or focus groups. Such information would help to assess the impact of trace selection errors. Another thing to keep in mind is the aspect of time, since existing articles may change, new articles may be added during the field observation, leading to unstable estimates. 
\smallskip
\smallskip

\subsubsection{Data Analysis}
  \hfill\\
  
\noindent
Finally, weekly flu rates are estimated using a Generalised Linear Model (GLM) on the article views. The authors' outcome variable is the proportion of positive age-weighted CDC ILI activity and the 35 predictor variables include the views on the 32 ILI-related articles, year and month. It is assumed that the GLM learns how ILI occurrences fluctuate based on the changes in the article views.

%\noindent
%\textbf{Unit of Analysis:} Assumed to be American Wikipedia readers 
\smallskip
\noindent
\textbf{Trace Measurement Errors:} Aggregated traces %(here aggregated views that correspond to our \textbf{unit of observation}) 
cannot be matched to users that produced them. It is, therefore, difficult to identify if multiple traces are referring to the same  user (multiple views of ILI-related Wikipedia pages made by the same person). As such, views of power users of Wikipedia are counted much more often than the average readers'.

\section{Related Work}\label{sec:related}
Our error framework for digital traces is inspired by insights from two disciplines: those that have been developed and refined for decades in (i) \textbf{survey methodology} as well as insights from the relatively new but rapidly developing field of (ii) \textbf{computational social science}, which has increasingly sought to understand the uses of newer forms data developed in the digital age. In this section, we discuss some of the main threads of research spanning these fields.

\textbf{Survey Methods.} The Total Survey Error Framework is an amalgamation of efforts to consolidate the different errors in the survey pipeline~\citep{groves2010total,biemer2008weighting}. Recently, researchers have tried to explore the efficacy of non-probability sampled surveys such as opt-in web surveys~\citep{goel2015non,goel2017online}, where they find that adjustment methods can improve estimates even if there are representation errors. On a similar vein,~\cite{kohler2019nonprobability} find that nonprobability samples can be effective in certain cases.~\cite{meng2018statistical} poses the question ``Which one should we trust more, a 5\% survey sample or an 80\% administrative dataset?" and introduces the concept of the `data defect index' to make surveys more comparable. Researchers have also attempted to integrate survey data and digital traces to analyze and uncover social scientific questions~\citep{stier2019integrating}.

\textbf{The Potentials of Digital Traces.} Web and social media data has been leveraged to `predict the future' in many areas such as politics, health, and economics~\citep{askitas2015internet,phillips2017using}. It is also considered as a means for learning about the present, e.g., for gaining insights into human behavior and opinions, such as using search queries to examine agenda-setting effects~\citep{ripberger2011capturing}, leveraging Instagram to detect drug use~\citep{yang2017tracking}, and measure consumer confidence through Twitter~\citep{pasek2018stability}. Acknowledging the potential advantages of digital trace data, the American Association for Public Opinion Research (AAPOR) has commissioned reports to understand the feasibility of using digital trace data for public opinion research~\citep{japec2015big,murphy2018social}.

\textbf{The Pitfalls of Digital Traces.} There is some important work aimed at uncovering errors that arise from using online data or, more generally, various kinds of digital trace data. Recently, researchers have studied the pitfalls related to political science research using digital trace data~\citep{gayo2012wanted,metaxas2011not,diaz2016online}, with ~\citet{jungherr2017digital,bender2021dangers}, especially focusing on issues of validity. Researchers have also analysed biases in digital traces arising due to demographic differences~\citep{pavalanathan2015confounds,olteanu2016characterizing}, platform effects~\citep{malik2016identifying} and data availability~\citep{morstatter2013sample,pfeffer2018tampering}. More generally, \citet{olteanu2016social}, provide a comprehensive overview of the errors and biases that could potentially affect studies based on digital behavioral data as well as outlining the errors in an idealized study framework, while \citet{tufekci2014big} outlines errors that can occur in Twitter-based studies. Recently~\citet{jungherr2017normalizing} calls for conceptualizing a measurement theory that may adequately account for the pitfalls of digital traces.    

\textbf{Addressing and Documenting the Pitfalls of Digital Traces. }Recently, researchers working with digital traces have used techniques typically used in surveys, to correct representation errors~\citep{zagheni2017leveraging,fatehkia2018using, Wang:2019:DIR:3308558.3313684}. Besides, addressing specific biases, researchers have also attempted to identify and document the different errors in digital traces as a first step to addressing them. In addition to an overview of biases in digital trace data, Olteanu et al. provide a framework using an idealized pipeline to enumerate various errors and how they may arise~\citep{olteanu2016social}. While highly comprehensive, they do not establish a connection with the Total Survey Error framework~\citep{groves2010total}, despite the similarities of errors that plague surveys as well. Ruths and Pfeffer prescribe actions that researchers can follow to reduces errors in social media data in two steps: data collection and methods~\citep{ruths2014social}. We extend this work in two ways: (1) by expanding our understanding of errors beyond social media platforms to other forms of digital trace data and (2) by taking a deep dive at more fine-grained steps to understand which design decision by a researcher contributes to what kind of error. 

 \citet{hsieh2017total} conceptualize the Total Twitter Error (TTE) Framework where they describe three kinds of errors in studying unstructured Twitter textual data, which can be mapped to survey errors: coverage error, query error and interpretation error. The authors also provide an empirical study of inferring attitudes towards political issues from tweets and limit their error framework to studies which follow a similar lifecycle. We aim to extend this framework to more diverse inference strategies, beyond textual analysis as well as for social media platforms other than Twitter and other forms of digital trace data. 
 
 The AAPOR report on public opinion assessment from `Big Data' sources (which includes large data sources other than digital trace data such as traffic and infrastructure data) describes a way to extend the TSE to such data, but cautions that a potential framework will have to account for errors that are specific to big data~\citep{japec2015big}. We restrict our error framework to digital trace data from the web which are collected based on a typical social sensing study pipeline, highlight where and how some of the new types of errors may arise, and how researchers may tackle them. Recent efforts to document issues in using digital trace data include ``Datasheets for Datasets'', proposed by~\cite{gebru2018datasheets} where datasets are accompanied with a datasheet that contains its motivation, composition, collection process, recommended uses as well as `Model cards' which document the use cases for machine learning models~\citep{mitchell2019model}. We propose a similar strategy to document digital trace based research designs to enable better communication, transparency, reproducibility, and reuse of said data.

Closest to our work is ~\citet{amaya2020total}, who apply the Total Error perspective to surveys and Big Data in general in their Total Error Framework (TEF). The TEF, however -- as all of the aforementioned TSE-inspired frameworks -- does not distinguish between representation and measurement errors, a differentiation which enables a more precise diagnosis of errors~\citep{groves2010total}. Moreover, the TED-On focuses on digital trace data of humans on the web in particular, as a counterpart to more general error frameworks. This allows for addressing the idiosyncrasies of digital platforms and new  methods, and putting forth concrete and clearly distinct errors, unambiguously linked to design steps in the research process. This regards, e.g.,  platform affordances, the need for complex machine-aided augmentation of users and traces, including unique errors not covered by the TSE. Lastly, combining digital traces and survey data is a promising avenue of making the best of both paradigms~\citep{stier2019integrating}. Beyond the work presented, we hope to extend the TED-On in order to address the ex-ante or ex-post linkage of both data sources, giving rise to new error types.

%Our TSE-inspired error framework leverages the common goal of surveys and digital traces to measure the attitudes, behaviours and characteristics of humans, and translates and adapts concepts developed in survey methodology for systematically documenting errors to that of digital trace based paradigm.
\section{Conclusion}\label{sec:conclusion}

The use of digital traces of human behavior, attitudes and opinion, %attitudes},
%\mr{and opinion -- correct? Almost any study using tweets as its data concerns opinion.}
especially web and social media data, is gaining traction in research communities, including the social sciences. %Application areas of this data include many forms of social sensing, e.g., public opinion research. 
In a growing number of scenarios, online digital traces are considered valuable complements to surveys. %, if not a less time-consuming and less costly alternative. \mr{you have not addressed cost and timeliness above so please either provide a relevant citation or delete this phrase.}
%It certainly allows scholars to study human behavior and attitudes at an unprecedented scale and the (inductive) study of how humans act and express themselves in novel digital contexts.
To make research on human behavior and opinions that are either based on online data, surveys, or both, more comparable based on shared vocabulary
%\mr{not entirely clear how TED makes the two data sources more comparable. Do you mean because it maps TED to TSE? Please be more explicit. } 
and to increase the reproducibility of online data-based research, it is important to describe potential error sources systematically. 

We have therefore introduced a framework for disentangling the various kinds of errors that may originate in the unsolicited nature of this data, in effects related to specific platforms, and due to the researchers' design choices in data collection, preprocessing and analysis. Drawing from concepts developed and refined in survey research, we provide suggestions for comparing these errors to their survey methodology counterpart based on the TSE, in order to (i) develop a shared vocabulary to enhance the dialogue among scientists from heterogeneous disciplines working in the area of computational social science, and (ii) to aid in pinpointing, documenting, and ultimately avoiding errors in research based on online digital traces. This (iii) might also lay the foundation for better research designs by calling attention to the distinct elements of the research pipeline that can be sources of error.

Just as survey methodology has benefited from understanding potential limitations in a systematic manner, our proposed framework can act as a set of recommendations for researchers on \textit{what} to reflect on when using online digital traces for studying social indicators. 
We recommend researchers not only use our framework to think about the various types of errors that may occur, but also use it to more systematically document limitations if they cannot be avoided. These design documents should be shared with research results where possible.
Using TED is the first step to improving estimates from human digital traces online since it allows us to transparently document errors. Future work will hopefully build on and extend the foundation laid by  TED-On to effectively quantify and mitigate errors inherent to digital trace data.

\textbf{Acknowledgments.} We thank the anonymous reviewers and the Special issue editors of POQ, especially Dr. Frederick G Conrad, Haiko Lietz, Sebastian Stier, Maria Zens, members of the GESIS CSS Department, Anna-Carolina Haensch, as well as participants of the Demography workshop at ICWSM'2019 for their helpful feedback and suggestions.

%In future work, we hope to tackle empirical digital trace data-based studies through the perspective of this error framework and attempt to quantify the errors and biases we encounter.

% CLAUDIA
% Some thoughts for follow up sections for the next version
% Application Examples of Framework
% Beside the Twitter and Reddit opinion mining examples, we could also add an example where the theoretical construct of interest is related with a network measure (e.g. power of an individual, homophily in a group). We could connect the theoretical examples with research that either is a positive or negative example. E.g. I found 2 nice papers that validate the meaning of social ties in FB in Twitter:
% https://dl.acm.org/citation.cfm?id=2380763
% https://dl.acm.org/citation.cfm?id=1518736

% Validation of real emotions versus emotional words on social media
% https://dl.acm.org/citation.cfm?id=2786473

% Macy-paper has lots of nice examples as well: https://www.annualreviews.org/doi/abs/10.1146/annurev-soc-071913-043145

% Finally I suggest that we also add some Recommendations like Howison

%\newpage
\footnotesize
\bibliographystyle{aaai}
\balance
\bibliography{main} 

\begin{thebibliography}{}

\bibitem[\protect\citeauthoryear{Alipourfard, Fennell, and
  Lerman}{2018}]{alipourfard2018can}
Alipourfard, N.; Fennell, P.~G.; and Lerman, K.
\newblock 2018.
\newblock Can you trust the trend?: Discovering simpson's paradoxes in social
  data.
\newblock In {\em Proceedings of the Eleventh ACM International Conference on
  Web Search and Data Mining},  19--27.
\newblock ACM.

\bibitem[\protect\citeauthoryear{Alzahrani \bgroup et al\mbox.\egroup
  }{2018}]{alzahrani2018finding}
Alzahrani, S.; Gore, C.; Salehi, A.; and Davulcu, H.
\newblock 2018.
\newblock Finding organizational accounts based on structural and behavioral
  factors on twitter.
\newblock In {\em International Conference on Social Computing,
  Behavioral-Cultural Modeling and Prediction and Behavior Representation in
  Modeling and Simulation},  164--175.
\newblock Springer.

\bibitem[\protect\citeauthoryear{Amaya, Biemer, and
  Kinyon}{2020}]{amaya2020total}
Amaya, A.; Biemer, P.~P.; and Kinyon, D.
\newblock 2020.
\newblock Total error in a big data world: Adapting the tse framework to big
  data.
\newblock {\em Journal of Survey Statistics and Methodology} 8(1):89--119.

\bibitem[\protect\citeauthoryear{An and Weber}{2015}]{an2015whom}
An, J., and Weber, I.
\newblock 2015.
\newblock Whom should we sense in “social sensing”-analyzing which users
  work best for social media now-casting.
\newblock {\em EPJ Data Science} 4(1):22.

\bibitem[\protect\citeauthoryear{Askitas and
  Zimmermann}{2015}]{askitas2015internet}
Askitas, N., and Zimmermann, K.~F.
\newblock 2015.
\newblock The internet as a data source for advancement in social sciences.
\newblock {\em International Journal of Manpower} 36(1):2--12.

\bibitem[\protect\citeauthoryear{Baker}{2017}]{bakerBigDataSurvey2017}
Baker, R.
\newblock 2017.
\newblock Big {{Data}}: {{A Survey Research Perspective}}.
\newblock In Biemer, P.~P.; {de Leeuw}, E.; Eckman, S.; Edwards, B.; Kreuter,
  F.; Lyberg, L.~E.; Tucker, N.~C.; and West, B.~T., eds., {\em Total {{Survey
  Error}} in {{Practice}}}. {Hoboken, NJ, USA}: {John Wiley \& Sons, Inc.}
\newblock  47--69.

\bibitem[\protect\citeauthoryear{Barber{\'a}}{2016}]{barbera2016less}
Barber{\'a}, P.
\newblock 2016.
\newblock Less is more? how demographic sample weights can improve public
  opinion estimates based on twitter data.
\newblock {\em Work Paper NYU}.

\bibitem[\protect\citeauthoryear{Bekafigo and
  McBride}{2013}]{bekafigo2013tweets}
Bekafigo, M.~A., and McBride, A.
\newblock 2013.
\newblock Who tweets about politics? political participation of twitter users
  during the 2011gubernatorial elections.
\newblock {\em Social Science Computer Review} 31(5):625--643.

\bibitem[\protect\citeauthoryear{Bender \bgroup et al\mbox.\egroup
  }{2021}]{bender2021dangers}
Bender, E.~M.; Gebru, T.; McMillan-Major, A.; and Shmitchell, S.
\newblock 2021.
\newblock On the dangers of stochastic parrots: Can language models be too big?
\newblock In {\em Proceedings of the 2021 ACM Conference on Fairness,
  Accountability, and Transparency},  610--623.

\bibitem[\protect\citeauthoryear{Biemer and Christ}{2008}]{biemer2008weighting}
Biemer, P., and Christ, S.
\newblock 2008.
\newblock Weighting survey data.
\newblock In {\em In de Leeuw ED, Hox JJ, Dillman DA, eds. International
  handbook of survey methodology}.
\newblock New York: Lawrence Erlbaum.

\bibitem[\protect\citeauthoryear{Biemer}{2010}]{biemer2010total}
Biemer, P.~P.
\newblock 2010.
\newblock Total survey error: Design, implementation, and evaluation.
\newblock {\em Public Opinion Quarterly} 74(5):817--848.

\bibitem[\protect\citeauthoryear{Borgatti, Carley, and
  Krackhardt}{2006}]{Borgatti2006}
Borgatti, S.; Carley, K.; and Krackhardt, D.
\newblock 2006.
\newblock Robustness of centrality measures under conditions of imperfect data.
\newblock {\em Social Networks} 28(1):124–136.

\bibitem[\protect\citeauthoryear{Bruns and Stieglitz}{2014}]{bruns2014twitter}
Bruns, A., and Stieglitz, S.
\newblock 2014.
\newblock Twitter data: what do they represent?
\newblock {\em it-Information Technology} 56(5):240--245.

\bibitem[\protect\citeauthoryear{Bruns and Weller}{2016}]{bruns2016twitter}
Bruns, A., and Weller, K.
\newblock 2016.
\newblock Twitter as a first draft of the present: and the challenges of
  preserving it for the future.
\newblock In {\em Proceedings of the 8th ACM Conference on Web Science},
  183--189.
\newblock ACM.

\bibitem[\protect\citeauthoryear{Buntain and Golbeck}{2015}]{buntain2015your}
Buntain, C., and Golbeck, J.
\newblock 2015.
\newblock This is your twitter on drugs: Any questions?
\newblock In {\em Proceedings of the 24th international conference on World
  Wide Web},  777--782.
\newblock ACM.

\bibitem[\protect\citeauthoryear{Buolamwini and
  Gebru}{2018}]{buolamwini2018gender}
Buolamwini, J., and Gebru, T.
\newblock 2018.
\newblock Gender shades: Intersectional accuracy disparities in commercial
  gender classification.
\newblock In {\em Conference on Fairness, Accountability and Transparency},
  77--91.

\bibitem[\protect\citeauthoryear{Cha \bgroup et al\mbox.\egroup
  }{2010}]{cha2010measuring}
Cha, M.; Haddadi, H.; Benevenuto, F.; and Gummadi, K.~P.
\newblock 2010.
\newblock Measuring user influence in twitter: The million follower fallacy.
\newblock In {\em fourth international AAAI conference on weblogs and social
  media}.

\bibitem[\protect\citeauthoryear{Chandrasekharan \bgroup et al\mbox.\egroup
  }{2017}]{chandrasekharan2017you}
Chandrasekharan, E.; Pavalanathan, U.; Srinivasan, A.; Glynn, A.; Eisenstein,
  J.; and Gilbert, E.
\newblock 2017.
\newblock You can't stay here: The efficacy of reddit's 2015 ban examined
  through hate speech.
\newblock {\em Proceedings of the ACM on Human-Computer Interaction}
  1(CSCW):31.

\bibitem[\protect\citeauthoryear{Cohen and Ruths}{2013}]{cohen2013classifying}
Cohen, R., and Ruths, D.
\newblock 2013.
\newblock Classifying political orientation on twitter: It’s not easy!
\newblock In {\em Seventh International AAAI Conference on Weblogs and Social
  Media}.

\bibitem[\protect\citeauthoryear{Conover \bgroup et al\mbox.\egroup
  }{2011}]{conover2011predicting}
Conover, M.~D.; Gon{\c{c}}alves, B.; Ratkiewicz, J.; Flammini, A.; and Menczer,
  F.
\newblock 2011.
\newblock Predicting the political alignment of twitter users.
\newblock In {\em 2011 IEEE third international conference on privacy,
  security, risk and trust and 2011 IEEE third international conference on
  social computing},  192--199.
\newblock IEEE.

\bibitem[\protect\citeauthoryear{Conrad \bgroup et al\mbox.\egroup
  }{2019}]{conrad2019social}
Conrad, F.~G.; Gagnon-Bartsch, J.~A.; Ferg, R.~A.; Schober, M.~F.; Pasek, J.;
  and Hou, E.
\newblock 2019.
\newblock Social media as an alternative to surveys of opinions about the
  economy.
\newblock {\em Social Science Computer Review}  0894439319875692.

\bibitem[\protect\citeauthoryear{Cornesse and
  Blom}{2020}]{cornesse2020response}
Cornesse, C., and Blom, A.~G.
\newblock 2020.
\newblock Response quality in nonprobability and probability-based online
  panels.
\newblock {\em Sociological Methods \& Research}  0049124120914940.

\bibitem[\protect\citeauthoryear{Cornesse \bgroup et al\mbox.\egroup
  }{2020}]{cornesse2020review}
Cornesse, C.; Blom, A.~G.; Dutwin, D.; Krosnick, J.~A.; De~Leeuw, E.~D.;
  Legleye, S.; Pasek, J.; Pennay, D.; Phillips, B.; Sakshaug, J.~W.; et~al.
\newblock 2020.
\newblock A review of conceptual approaches and empirical evidence on
  probability and nonprobability sample survey research.
\newblock {\em Journal of Survey Statistics and Methodology} 8(1):4--36.

\bibitem[\protect\citeauthoryear{Costenbader and
  Valente}{2003}]{costenbader2003stability}
Costenbader, E., and Valente, T.~W.
\newblock 2003.
\newblock The stability of centrality measures when networks are sampled.
\newblock {\em Social networks} 25(4):283--307.

\bibitem[\protect\citeauthoryear{Culotta}{2014}]{culotta2014reducing}
Culotta, A.
\newblock 2014.
\newblock Reducing sampling bias in social media data for county health
  inference.
\newblock In {\em Joint Statistical Meetings Proceedings},  1--12.

\bibitem[\protect\citeauthoryear{De~Cristofaro \bgroup et al\mbox.\egroup
  }{2018}]{de2018lobo}
De~Cristofaro, E.; Kourtellis, N.; Leontiadis, I.; Stringhini, G.; Zhou, S.;
  et~al.
\newblock 2018.
\newblock Lobo: Evaluation of generalization deficiencies in twitter bot
  classifiers.
\newblock In {\em Proceedings of the 34th Annual Computer Security Applications
  Conference},  137--146.
\newblock ACM.

\bibitem[\protect\citeauthoryear{Demartini}{2007}]{demartini2007finding}
Demartini, G.
\newblock 2007.
\newblock Finding experts using wikipedia.
\newblock In {\em Proceedings of the 2nd International Conference on Finding
  Experts on the Web with Semantics-Volume 290},  33--41.
\newblock Citeseer.

\bibitem[\protect\citeauthoryear{Diaz \bgroup et al\mbox.\egroup
  }{2016}]{diaz2016online}
Diaz, F.; Gamon, M.; Hofman, J.~M.; K{\i}c{\i}man, E.; and Rothschild, D.
\newblock 2016.
\newblock Online and social media data as an imperfect continuous panel survey.
\newblock {\em PloS one} 11(1):e0145406.

\bibitem[\protect\citeauthoryear{DiMaggio \bgroup et al\mbox.\egroup
  }{2001}]{DiMaggio2011}
DiMaggio, P.; Hargittai, E.; Neuman, W.~R.; and Robinson, J.~P.
\newblock 2001.
\newblock Social implications of the internet.
\newblock {\em Annual Review of Sociology} 27(1):307--336.

\bibitem[\protect\citeauthoryear{Dimitrov \bgroup et al\mbox.\egroup
  }{2019}]{dimitrov2019different}
Dimitrov, D.; Lemmerich, F.; Fl{\"o}ck, F.; and Strohmaier, M.
\newblock 2019.
\newblock Different topic, different trafic: How search and navigation
  interplay on wikipedia.
\newblock {\em The Journal of Web Science} 1.

\bibitem[\protect\citeauthoryear{Duggan and Smith}{2013}]{duggan20136}
Duggan, M., and Smith, A.
\newblock 2013.
\newblock 6\% of online adults are reddit users.
\newblock {\em Pew Internet \& American Life Project} 3:1--10.

\bibitem[\protect\citeauthoryear{Eckman and
  Kreuter}{2017}]{eckman2017undercoverage}
Eckman, S., and Kreuter, F.
\newblock 2017.
\newblock The undercoverage-nonresponse trade-off.
\newblock {\em Total Survey Error in Practice}  95--113.

\bibitem[\protect\citeauthoryear{Fatehkia, Kashyap, and
  Weber}{2018}]{fatehkia2018using}
Fatehkia, M.; Kashyap, R.; and Weber, I.
\newblock 2018.
\newblock Using facebook ad data to track the global digital gender gap.
\newblock {\em World Development} 107:189--209.

\bibitem[\protect\citeauthoryear{Fiesler and
  Proferes}{2018}]{fiesler2018participant}
Fiesler, C., and Proferes, N.
\newblock 2018.
\newblock “participant” perceptions of twitter research ethics.
\newblock {\em Social Media \& Society} 4(1):2056305118763366.

\bibitem[\protect\citeauthoryear{Fischer and
  Budescu}{1995}]{fischer1995desirability}
Fischer, I., and Budescu, D.~V.
\newblock 1995.
\newblock Desirability and hindsight biases in predicting results of a
  multi-party election.

\bibitem[\protect\citeauthoryear{Frankze \bgroup et al\mbox.\egroup
  }{2019}]{AOIR2019Internet}
Frankze, A.~S.; Bechmann, A.; Zimmer, M.; and Ess, C.~M.
\newblock 2019.
\newblock Internet research: Ethical guidelines 3.0.

\bibitem[\protect\citeauthoryear{Galaskiewicz}{1991}]{Galaskiewicz1991}
Galaskiewicz, J.
\newblock 1991.
\newblock {Estimating point centrality using different network sampling
  techniques}.
\newblock {\em Social Networks} 13(4):347--386.

\bibitem[\protect\citeauthoryear{Garcia \bgroup et al\mbox.\egroup
  }{2018}]{garcia2018collective}
Garcia, D.; Goel, M.; Agrawal, A.~K.; and Kumaraguru, P.
\newblock 2018.
\newblock Collective aspects of privacy in the twitter social network.
\newblock {\em EPJ Data Science} 7(1):3.

\bibitem[\protect\citeauthoryear{Gayo-Avello}{2012}]{gayo2012wanted}
Gayo-Avello, D.
\newblock 2012.
\newblock ``i wanted to predict elections with twitter and all i got was this
  lousy paper"--a balanced survey on election prediction using twitter data.
\newblock {\em arXiv preprint arXiv:1204.6441}.

\bibitem[\protect\citeauthoryear{Gebru \bgroup et al\mbox.\egroup
  }{2018}]{gebru2018datasheets}
Gebru, T.; Morgenstern, J.; Vecchione, B.; Vaughan, J.~W.; Wallach, H.;
  Daume{\'e}~III, H.; and Crawford, K.
\newblock 2018.
\newblock Datasheets for datasets.
\newblock {\em arXiv preprint arXiv:1803.09010}.

\bibitem[\protect\citeauthoryear{Gligori{\'c}, Anderson, and
  West}{2018}]{gligoric2018constraints}
Gligori{\'c}, K.; Anderson, A.; and West, R.
\newblock 2018.
\newblock How constraints affect content: The case of twitter’s switch from
  140 to 280 characters.
\newblock In {\em Twelfth International AAAI Conference on Web and Social
  Media}.

\bibitem[\protect\citeauthoryear{Goel, Obeng, and
  Rothschild}{2015}]{goel2015non}
Goel, S.; Obeng, A.; and Rothschild, D.
\newblock 2015.
\newblock Non-representative surveys: Fast, cheap, and mostly accurate.
\newblock {\em Working Paper}.

\bibitem[\protect\citeauthoryear{Goel, Obeng, and
  Rothschild}{2017}]{goel2017online}
Goel, S.; Obeng, A.; and Rothschild, D.
\newblock 2017.
\newblock Online, opt-in surveys: Fast and cheap, but are they accurate?
\newblock {\em Working Paper, Stanford University, Stanford, CA}.

\bibitem[\protect\citeauthoryear{Gonz{\'a}lez-Bail{\'o}n and
  Paltoglou}{2015}]{gonzalez2015signals}
Gonz{\'a}lez-Bail{\'o}n, S., and Paltoglou, G.
\newblock 2015.
\newblock Signals of public opinion in online communication: A comparison of
  methods and data sources.
\newblock {\em The ANNALS of the American Academy of Political and Social
  Science} 659(1):95--107.

\bibitem[\protect\citeauthoryear{Gonz{\'a}lez-Bail{\'o}n \bgroup et
  al\mbox.\egroup }{2014}]{gonzalez2014assessing}
Gonz{\'a}lez-Bail{\'o}n, S.; Wang, N.; Rivero, A.; Borge-Holthoefer, J.; and
  Moreno, Y.
\newblock 2014.
\newblock Assessing the bias in samples of large online networks.
\newblock {\em Social Networks} 38:16--27.

\bibitem[\protect\citeauthoryear{Groves and Lyberg}{2010}]{groves2010total}
Groves, R.~M., and Lyberg, L.
\newblock 2010.
\newblock Total survey error: Past, present, and future.
\newblock {\em Public opinion quarterly} 74(5):849--879.

\bibitem[\protect\citeauthoryear{Groves \bgroup et al\mbox.\egroup
  }{2009}]{groves2011survey}
Groves, R.~M.; Fowler~Jr, F.~J.; Couper, M.~P.; Lepkowski, J.~M.; Singer, E.;
  and Tourangeau, R.
\newblock 2009.
\newblock {\em Survey methodology}, volume 561.
\newblock John Wiley \& Sons.

\bibitem[\protect\citeauthoryear{Groves}{2011}]{groves_three_2011}
Groves, R.~M.
\newblock 2011.
\newblock Three {Eras} of {Survey} {Research}.
\newblock {\em Public Opinion Quarterly} 75(5):861--871.

\bibitem[\protect\citeauthoryear{Hamidi, Scheuerman, and
  Branham}{2018}]{hamidi2018gender}
Hamidi, F.; Scheuerman, M.~K.; and Branham, S.~M.
\newblock 2018.
\newblock Gender recognition or gender reductionism? the social implications of
  embedded gender recognition systems.
\newblock In {\em Proceedings of the 2018 chi conference on human factors in
  computing systems},  1--13.

\bibitem[\protect\citeauthoryear{Hamilton \bgroup et al\mbox.\egroup
  }{2016}]{hamilton2016inducing}
Hamilton, W.~L.; Clark, K.; Leskovec, J.; and Jurafsky, D.
\newblock 2016.
\newblock Inducing domain-specific sentiment lexicons from unlabeled corpora.
\newblock In {\em Proceedings of the Conference on Empirical Methods in Natural
  Language Processing. Conference on Empirical Methods in Natural Language
  Processing}, volume 2016,  595.
\newblock NIH Public Access.

\bibitem[\protect\citeauthoryear{He and Rothschild}{2016}]{he2016selection}
He, R., and Rothschild, D.
\newblock 2016.
\newblock Selection bias in documenting online conversations.
\newblock {\em Working paper}.

\bibitem[\protect\citeauthoryear{Howison, Wiggins, and
  Crowston}{2011}]{Howison2011}
Howison, J.; Wiggins, A.; and Crowston, K.
\newblock 2011.
\newblock Validity issues in the use of social network analysis with digital
  trace data.
\newblock {\em Journal of the Association for Information Systems} 12(12).

\bibitem[\protect\citeauthoryear{Hsieh and Murphy}{2017}]{hsieh2017total}
Hsieh, Y.~P., and Murphy, J.
\newblock 2017.
\newblock Total twitter error.
\newblock {\em Total Survey Error in Practice}  23--46.

\bibitem[\protect\citeauthoryear{Jacobs and
  Wallach}{2019}]{jacobs2019measurement}
Jacobs, A.~Z., and Wallach, H.
\newblock 2019.
\newblock Measurement and fairness.
\newblock {\em arXiv preprint arXiv:1912.05511}.

\bibitem[\protect\citeauthoryear{Japec \bgroup et al\mbox.\egroup
  }{2015}]{japec2015big}
Japec, L.; Kreuter, F.; Berg, M.; Biemer, P.; Decker, P.; Lampe, C.; Lane, J.;
  O’Neil, C.; and Usher, A.
\newblock 2015.
\newblock Big data in survey research: Aapor task force report.
\newblock {\em Public Opinion Quarterly} 79(4):839--880.

\bibitem[\protect\citeauthoryear{Jha and Mamidi}{2017}]{jha2017does}
Jha, A., and Mamidi, R.
\newblock 2017.
\newblock When does a compliment become sexist? analysis and classification of
  ambivalent sexism using twitter data.
\newblock In {\em Proceedings of the second workshop on NLP and computational
  social science},  7--16.

\bibitem[\protect\citeauthoryear{Johnson, Safadi, and
  Faraj}{2015}]{johnson2015emergence}
Johnson, S.~L.; Safadi, H.; and Faraj, S.
\newblock 2015.
\newblock The emergence of online community leadership.
\newblock {\em Information Systems Research} 26(1):165--187.

\bibitem[\protect\citeauthoryear{Joye \bgroup et al\mbox.\egroup
  }{2016}]{wolf_survey_2016}
Joye, D.; Wolf, C.; Smith, T.~W.; and Fu, Y.-c.
\newblock 2016.
\newblock Survey {Methodology}: {Challenges} and {Principles}.
\newblock In {\em The {SAGE} {Handbook} of {Survey} {Methodology}}. 1 Oliver's
  Yard, 55 City Road London EC1Y 1SP: SAGE Publications Ltd.
\newblock  3--15.

\bibitem[\protect\citeauthoryear{Jungherr \bgroup et al\mbox.\egroup
  }{2017}]{jungherr2017digital}
Jungherr, A.; Schoen, H.; Posegga, O.; and J{\"u}rgens, P.
\newblock 2017.
\newblock Digital trace data in the study of public opinion: An indicator of
  attention toward politics rather than political support.
\newblock {\em Social Science Computer Review} 35(3):336--356.

\bibitem[\protect\citeauthoryear{Jungherr}{2017}]{jungherr2017normalizing}
Jungherr, A.
\newblock 2017.
\newblock Normalizing digital trace data.
\newblock {\em Digital discussions: How big data informs political
  communication}.

\bibitem[\protect\citeauthoryear{Karimi \bgroup et al\mbox.\egroup
  }{2016}]{Karimi2016}
Karimi, F.; Wagner, C.; Lemmerich, F.; Jadidi, M.; and Strohmaier, M.
\newblock 2016.
\newblock Inferring gender from names on the web: A comparative evaluation of
  gender detection methods.
\newblock In {\em Proceedings of the 25th International Conference Companion on
  World Wide Web}, WWW '16 Companion,  53--54.
\newblock International World Wide Web Conferences Steering Committee.

\bibitem[\protect\citeauthoryear{Keyes}{2018}]{keyes2018misgendering}
Keyes, O.
\newblock 2018.
\newblock The misgendering machines: Trans/hci implications of automatic gender
  recognition.
\newblock {\em Proceedings of the ACM on human-computer interaction}
  2(CSCW):1--22.

\bibitem[\protect\citeauthoryear{Khan and Fu}{2021}]{khan2021one}
Khan, Z., and Fu, Y.
\newblock 2021.
\newblock One label, one billion faces: Usage and consistency of racial
  categories in computer vision.
\newblock In {\em Proceedings of the 2021 ACM Conference on Fairness,
  Accountability, and Transparency},  587--597.

\bibitem[\protect\citeauthoryear{Kohler, Kreuter, and
  Stuart}{2019}]{kohler2019nonprobability}
Kohler, U.; Kreuter, F.; and Stuart, E.~A.
\newblock 2019.
\newblock Nonprobability sampling and causal analysis.
\newblock {\em Annual review of statistics and its application} 6:149--172.

\bibitem[\protect\citeauthoryear{Kohler}{2019}]{kohler_possible_2019}
Kohler, U.
\newblock 2019.
\newblock Possible {Uses} of {Nonprobability} {Sampling} for the {Social}
  {Sciences}.
\newblock {\em Survey Methods: Insights from the Field}.

\bibitem[\protect\citeauthoryear{Kossinets}{2006}]{Kossinets2006}
Kossinets, G.
\newblock 2006.
\newblock Effects of missing data in social networks.
\newblock {\em Social Networks} 28:247--268.

\bibitem[\protect\citeauthoryear{Kumar and Shah}{2018}]{kumar2018false}
Kumar, S., and Shah, N.
\newblock 2018.
\newblock False information on web and social media: A survey.
\newblock {\em arXiv preprint arXiv:1804.08559}.

\bibitem[\protect\citeauthoryear{Lamprecht \bgroup et al\mbox.\egroup
  }{2017}]{lamprecht2017structure}
Lamprecht, D.; Lerman, K.; Helic, D.; and Strohmaier, M.
\newblock 2017.
\newblock How the structure of wikipedia articles influences user navigation.
\newblock {\em New Review of Hypermedia and Multimedia} 23(1):29--50.

\bibitem[\protect\citeauthoryear{Lazer \bgroup et al\mbox.\egroup
  }{2009}]{lazer2009social}
Lazer, D.; Pentland, A.; Adamic, L.; Aral, S.; Barabasi, A.-L.; Brewer, D.;
  Christakis, N.; Contractor, N.; Fowler, J.; Gutmann, M.; et~al.
\newblock 2009.
\newblock Social science. computational social science.
\newblock {\em Science (New York, NY)} 323(5915):721--723.

\bibitem[\protect\citeauthoryear{Lazer}{2015}]{lazer2015issues}
Lazer, D.
\newblock 2015.
\newblock Issues of construct validity and reliability in massive, passive data
  collections.
\newblock {\em The City Papers: An Essay Collection from The Decent City
  Initiative}.

\bibitem[\protect\citeauthoryear{Lee and Pfeffer}{2015}]{Lee2015}
Lee, J., and Pfeffer, J.
\newblock 2015.
\newblock Estimating centrality statistics for complete and sampled networks:
  Some approaches and complications.
\newblock In {\em 48th Hawaii International Conference on System Sciences,
  {HICSS} 2015, Kauai, Hawaii, USA, January 5-8, 2015},  1686--1695.

\bibitem[\protect\citeauthoryear{Lerman}{2018}]{lerman2018computational}
Lerman, K.
\newblock 2018.
\newblock Computational social scientist beware: Simpson’s paradox in
  behavioral data.
\newblock {\em Journal of Computational Social Science} 1(1):49--58.

\bibitem[\protect\citeauthoryear{Linder}{2017}]{linder2017improved}
Linder, F.
\newblock 2017.
\newblock Improved data collection from online sources using query expansion
  and active learning.
\newblock {\em Available at SSRN 3026393}.

\bibitem[\protect\citeauthoryear{Locker}{1993}]{locker1993effects}
Locker, D.
\newblock 1993.
\newblock Effects of non-response on estimates derived from an oral health
  survey of older adults.
\newblock {\em Community dentistry and oral epidemiology} 21(2):108--113.

\bibitem[\protect\citeauthoryear{Malik and
  Pfeffer}{2016}]{malik2016identifying}
Malik, M.~M., and Pfeffer, J.
\newblock 2016.
\newblock Identifying platform effects in social media data.
\newblock In {\em Tenth International AAAI Conference on Web and Social Media}.

\bibitem[\protect\citeauthoryear{McCormick \bgroup et al\mbox.\egroup
  }{2017}]{mccormick2017using}
McCormick, T.~H.; Lee, H.; Cesare, N.; Shojaie, A.; and Spiro, E.~S.
\newblock 2017.
\newblock Using twitter for demographic and social science research: Tools for
  data collection and processing.
\newblock {\em Sociological methods \& research} 46(3):390--421.

\bibitem[\protect\citeauthoryear{McCorriston, Jurgens, and
  Ruths}{2015}]{mccorriston2015organizations}
McCorriston, J.; Jurgens, D.; and Ruths, D.
\newblock 2015.
\newblock Organizations are users too: Characterizing and detecting the
  presence of organizations on twitter.
\newblock In {\em Ninth International AAAI Conference on Web and Social Media}.

\bibitem[\protect\citeauthoryear{McIver and
  Brownstein}{2014}]{mciver2014wikipedia}
McIver, D.~J., and Brownstein, J.~S.
\newblock 2014.
\newblock Wikipedia usage estimates prevalence of influenza-like illness in the
  united states in near real-time.
\newblock {\em PLoS computational biology} 10(4):e1003581.

\bibitem[\protect\citeauthoryear{McMahon, Johnson, and
  Hecht}{2017}]{mcmahon2017substantial}
McMahon, C.; Johnson, I.; and Hecht, B.
\newblock 2017.
\newblock The substantial interdependence of wikipedia and google: A case study
  on the relationship between peer production communities and information
  technologies.
\newblock In {\em Eleventh International AAAI Conference on Web and Social
  Media}.

\bibitem[\protect\citeauthoryear{Meng}{2018}]{meng2018statistical}
Meng, X.-L.
\newblock 2018.
\newblock Statistical paradises and paradoxes in big data (i): Law of large
  populations, big data paradox, and the 2016 us presidential election.
\newblock {\em The Annals of Applied Statistics} 12(2):685--726.

\bibitem[\protect\citeauthoryear{Metaxas, Mustafaraj, and
  Gayo-Avello}{2011}]{metaxas2011not}
Metaxas, P.~T.; Mustafaraj, E.; and Gayo-Avello, D.
\newblock 2011.
\newblock How (not) to predict elections.
\newblock In {\em 2011 IEEE Third International Conference on Privacy,
  Security, Risk and Trust and 2011 IEEE Third International Conference on
  Social Computing},  165--171.
\newblock IEEE.

\bibitem[\protect\citeauthoryear{Mislove \bgroup et al\mbox.\egroup
  }{2011}]{mislove2011understanding}
Mislove, A.; Lehmann, S.; Ahn, Y.-Y.; Onnela, J.-P.; and Rosenquist, J.~N.
\newblock 2011.
\newblock Understanding the demographics of twitter users.
\newblock In {\em Fifth international AAAI conference on weblogs and social
  media}.

\bibitem[\protect\citeauthoryear{Mitchell \bgroup et al\mbox.\egroup
  }{2019}]{mitchell2019model}
Mitchell, M.; Wu, S.; Zaldivar, A.; Barnes, P.; Vasserman, L.; Hutchinson, B.;
  Spitzer, E.; Raji, I.~D.; and Gebru, T.
\newblock 2019.
\newblock Model cards for model reporting.
\newblock In {\em Proceedings of the Conference on Fairness, Accountability,
  and Transparency},  220--229.
\newblock ACM.

\bibitem[\protect\citeauthoryear{Mittelstadt \bgroup et al\mbox.\egroup
  }{2016}]{mittelstadt2016ethics}
Mittelstadt, B.~D.; Allo, P.; Taddeo, M.; Wachter, S.; and Floridi, L.
\newblock 2016.
\newblock The ethics of algorithms: Mapping the debate.
\newblock {\em Big Data \& Society} 3(2):2053951716679679.

\bibitem[\protect\citeauthoryear{Morstatter \bgroup et al\mbox.\egroup
  }{2013}]{morstatter2013sample}
Morstatter, F.; Pfeffer, J.; Liu, H.; and Carley, K.~M.
\newblock 2013.
\newblock Is the sample good enough? comparing data from twitter's streaming
  api with twitter's firehose.
\newblock In {\em Seventh international AAAI conference on weblogs and social
  media}.

\bibitem[\protect\citeauthoryear{Murphy and others}{2018}]{murphy2018social}
Murphy, J., et~al.
\newblock 2018.
\newblock Social media in public opinion research: Report of the aapor task
  force on emerging technologies in public opinion research. american
  association for public opinion research. 2014.

\bibitem[\protect\citeauthoryear{O'Connor \bgroup et al\mbox.\egroup
  }{2010}]{o2010tweets}
O'Connor, B.; Balasubramanyan, R.; Routledge, B.~R.; and Smith, N.~A.
\newblock 2010.
\newblock From tweets to polls: Linking text sentiment to public opinion time
  series.
\newblock In {\em Fourth International AAAI Conference on Weblogs and Social
  Media}.

\bibitem[\protect\citeauthoryear{Olteanu \bgroup et al\mbox.\egroup
  }{2019}]{olteanu2016social}
Olteanu, A.; Castillo, C.; Diaz, F.; and Kiciman, E.
\newblock 2019.
\newblock Social data: Biases, methodological pitfalls, and ethical boundaries.
\newblock {\em Frontiers in Big Data} 2:13.

\bibitem[\protect\citeauthoryear{Olteanu, Weber, and
  Gatica-Perez}{2016}]{olteanu2016characterizing}
Olteanu, A.; Weber, I.; and Gatica-Perez, D.
\newblock 2016.
\newblock Characterizing the demographics behind the\# blacklivesmatter
  movement.
\newblock In {\em 2016 AAAI Spring Symposium Series}.

\bibitem[\protect\citeauthoryear{Pasek \bgroup et al\mbox.\egroup
  }{2018}]{pasek2018stability}
Pasek, J.; Yan, H.~Y.; Conrad, F.~G.; Newport, F.; and Marken, S.
\newblock 2018.
\newblock The stability of economic correlations over time: Identifying
  conditions under which survey tracking polls and twitter sentiment yield
  similar conclusions.
\newblock {\em Public Opinion Quarterly} 82(3):470--492.

\bibitem[\protect\citeauthoryear{Pasek \bgroup et al\mbox.\egroup
  }{2019}]{pasek2019s}
Pasek, J.; McClain, C.~A.; Newport, F.; and Marken, S.
\newblock 2019.
\newblock Who’s tweeting about the president? what big survey data can tell
  us about digital traces?
\newblock {\em Social Science Computer Review}  0894439318822007.

\bibitem[\protect\citeauthoryear{Pavalanathan and
  Eisenstein}{2015}]{pavalanathan2015confounds}
Pavalanathan, U., and Eisenstein, J.
\newblock 2015.
\newblock Confounds and consequences in geotagged twitter data.
\newblock {\em arXiv preprint arXiv:1506.02275}.

\bibitem[\protect\citeauthoryear{Pfeffer, Mayer, and
  Morstatter}{2018}]{pfeffer2018tampering}
Pfeffer, J.; Mayer, K.; and Morstatter, F.
\newblock 2018.
\newblock Tampering with twitter’s sample api.
\newblock {\em EPJ Data Science} 7(1):50.

\bibitem[\protect\citeauthoryear{Phillips \bgroup et al\mbox.\egroup
  }{2017}]{phillips2017using}
Phillips, L.; Dowling, C.; Shaffer, K.; Hodas, N.; and Volkova, S.
\newblock 2017.
\newblock Using social media to predict the future: a systematic literature
  review.
\newblock {\em arXiv preprint arXiv:1706.06134}.

\bibitem[\protect\citeauthoryear{Puschmann and
  Powell}{2018}]{puschmann2018turning}
Puschmann, C., and Powell, A.
\newblock 2018.
\newblock Turning words into consumer preferences: How sentiment analysis is
  framed in research and the news media.
\newblock {\em Social Media+ Society} 4(3):2056305118797724.

\bibitem[\protect\citeauthoryear{Rao \bgroup et al\mbox.\egroup
  }{2010}]{rao2010classifying}
Rao, D.; Yarowsky, D.; Shreevats, A.; and Gupta, M.
\newblock 2010.
\newblock Classifying latent user attributes in twitter.
\newblock In {\em Proceedings of the 2nd international workshop on Search and
  mining user-generated contents},  37--44.
\newblock ACM.

\bibitem[\protect\citeauthoryear{Ribeiro \bgroup et al\mbox.\egroup
  }{2020}]{ribeiro2020does}
Ribeiro, M.~H.; Jhaver, S.; Zannettou, S.; Blackburn, J.; De~Cristofaro, E.;
  Stringhini, G.; and West, R.
\newblock 2020.
\newblock Does platform migration compromise content moderation? evidence from
  r/the\_donald and r/incels.
\newblock {\em arXiv preprint arXiv:2010.10397}.

\bibitem[\protect\citeauthoryear{Riederer \bgroup et al\mbox.\egroup
  }{2015}]{riederer2015don}
Riederer, C.~J.; Zimmeck, S.; Phanord, C.; Chaintreau, A.; and Bellovin, S.~M.
\newblock 2015.
\newblock I don't have a photograph, but you can have my footprints.: Revealing
  the demographics of location data.
\newblock In {\em Proceedings of the 2015 ACM on Conference on Online Social
  Networks},  185--195.
\newblock ACM.

\bibitem[\protect\citeauthoryear{Ripberger}{2011}]{ripberger2011capturing}
Ripberger, J.~T.
\newblock 2011.
\newblock Capturing curiosity: Using internet search trends to measure public
  attentiveness.
\newblock {\em Policy Studies Journal} 39(2):239--259.

\bibitem[\protect\citeauthoryear{Ruiz, Hristidis, and
  Ipeirotis}{2014}]{ruiz2014efficient}
Ruiz, E.~J.; Hristidis, V.; and Ipeirotis, P.~G.
\newblock 2014.
\newblock Efficient filtering on hidden document streams.
\newblock In {\em Eighth International AAAI Conference on Weblogs and Social
  Media}.

\bibitem[\protect\citeauthoryear{Ruths and Pfeffer}{2014}]{ruths2014social}
Ruths, D., and Pfeffer, J.
\newblock 2014.
\newblock Social media for large studies of behavior.
\newblock {\em Science} 346(6213):1063--1064.

\bibitem[\protect\citeauthoryear{Sakaki, Okazaki, and
  Matsuo}{2010}]{sakaki2010earthquake}
Sakaki, T.; Okazaki, M.; and Matsuo, Y.
\newblock 2010.
\newblock Earthquake shakes twitter users: real-time event detection by social
  sensors.
\newblock In {\em Proceedings of the 19th international conference on World
  wide web},  851--860.
\newblock ACM.

\bibitem[\protect\citeauthoryear{Salganik}{2017}]{salganik2017bit}
Salganik, M.~J.
\newblock 2017.
\newblock {\em Bit by bit: social research in the digital age}.
\newblock Princeton University Press.

\bibitem[\protect\citeauthoryear{Samory \bgroup et al\mbox.\egroup
  }{2021}]{samory_call_2021}
Samory, M.; Sen, I.; Kohne, J.; Floeck, F.; and Wagner, C.
\newblock 2021.
\newblock Call me sexist, but..: {Revisiting} {Sexism} {Detection} {Using}
  {Psychological} {Scales} and {Adversarial} {Samples}.
\newblock arXiv: 2004.12764.

\bibitem[\protect\citeauthoryear{Sap \bgroup et al\mbox.\egroup
  }{2014}]{sap2014developing}
Sap, M.; Park, G.; Eichstaedt, J.; Kern, M.; Stillwell, D.; Kosinski, M.;
  Ungar, L.; and Schwartz, H.~A.
\newblock 2014.
\newblock Developing age and gender predictive lexica over social media.
\newblock In {\em Proceedings of the 2014 Conference on Empirical Methods in
  Natural Language Processing (EMNLP)},  1146--1151.

\bibitem[\protect\citeauthoryear{Scheuerman, Paul, and
  Brubaker}{2019}]{scheuerman2019computers}
Scheuerman, M.~K.; Paul, J.~M.; and Brubaker, J.~R.
\newblock 2019.
\newblock How computers see gender: An evaluation of gender classification in
  commercial facial analysis services.
\newblock {\em Proceedings of the ACM on Human-Computer Interaction}
  3(CSCW):1--33.

\bibitem[\protect\citeauthoryear{Schnell, Noack, and
  Torregroza}{2017}]{schnell2017differences}
Schnell, R.; Noack, M.; and Torregroza, S.
\newblock 2017.
\newblock Differences in general health of internet users and non-users and
  implications for the use of web surveys.
\newblock In {\em Survey Research Methods}, volume~11,  105--123.

\bibitem[\protect\citeauthoryear{Schober \bgroup et al\mbox.\egroup
  }{2016}]{schober2016social}
Schober, M.~F.; Pasek, J.; Guggenheim, L.; Lampe, C.; and Conrad, F.~G.
\newblock 2016.
\newblock Social media analyses for social measurement.
\newblock {\em Public opinion quarterly} 80(1):180--211.

\bibitem[\protect\citeauthoryear{Sen, Fl{\"o}ck, and
  Wagner}{2020}]{sen2020reliability}
Sen, I.; Fl{\"o}ck, F.; and Wagner, C.
\newblock 2020.
\newblock On the reliability and validity of detecting approval of political
  actors in tweets.
\newblock In {\em Proceedings of the 2020 Conference on Empirical Methods in
  Natural Language Processing (EMNLP)},  1413--1426.

\bibitem[\protect\citeauthoryear{Smith, Anderson, and
  others}{2018}]{smith2018social}
Smith, A.; Anderson, M.; et~al.
\newblock 2018.
\newblock Social media use in 2018.
\newblock {\em Pew research center} 1:1--4.

\bibitem[\protect\citeauthoryear{Stier \bgroup et al\mbox.\egroup
  }{2018}]{stier2018systematically}
Stier, S.; Bleier, A.; Bonart, M.; M{\"o}rsheim, F.; Bohlouli, M.;
  Nizhegorodov, M.; Posch, L.; Maier, J.; Rothmund, T.; and Staab, S.
\newblock 2018.
\newblock Systematically monitoring social media: The case of the german
  federal election 2017.
\newblock {\em arXiv preprint arXiv:1804.02888}.

\bibitem[\protect\citeauthoryear{Stier \bgroup et al\mbox.\egroup
  }{2019}]{stier2019integrating}
Stier, S.; Breuer, J.; Siegers, P.; and Thorson, K.
\newblock 2019.
\newblock Integrating survey data and digital trace data: key issues in
  developing an emerging field.

\bibitem[\protect\citeauthoryear{Straub, Boudreau, and
  Gefen}{2004}]{straub2004validation}
Straub, D.; Boudreau, M.-C.; and Gefen, D.
\newblock 2004.
\newblock Validation guidelines for is positivist research.
\newblock {\em Communications of the Association for Information systems}
  13(1):24.

\bibitem[\protect\citeauthoryear{Tufekci}{2014}]{tufekci2014big}
Tufekci, Z.
\newblock 2014.
\newblock Big questions for social media big data: Representativeness, validity
  and other methodological pitfalls.
\newblock In {\em Eighth International AAAI Conference on Weblogs and Social
  Media}.

\bibitem[\protect\citeauthoryear{Tumasjan \bgroup et al\mbox.\egroup
  }{2010}]{tumasjan2010predicting}
Tumasjan, A.; Sprenger, T.~O.; Sandner, P.~G.; and Welpe, I.~M.
\newblock 2010.
\newblock Predicting elections with twitter: What 140 characters reveal about
  political sentiment.
\newblock In {\em Fourth international AAAI conference on weblogs and social
  media}.

\bibitem[\protect\citeauthoryear{Wagner \bgroup et al\mbox.\egroup
  }{2017}]{wagner2017sampling}
Wagner, C.; Singer, P.; Karimi, F.; Pfeffer, J.; and Strohmaier, M.
\newblock 2017.
\newblock Sampling from social networks with attributes.
\newblock In {\em Proceedings of the 26th International Conference on World
  Wide Web},  1181--1190.
\newblock International World Wide Web Conferences Steering Committee.

\bibitem[\protect\citeauthoryear{Wang \bgroup et al\mbox.\egroup
  }{2012}]{wang2012measurement}
Wang, D.~J.; Shi, X.; McFarland, D.~A.; and Leskovec, J.
\newblock 2012.
\newblock Measurement error in network data: A re-classification.
\newblock {\em Social Networks} 34(4):396--409.

\bibitem[\protect\citeauthoryear{Wang \bgroup et al\mbox.\egroup
  }{2019}]{Wang:2019:DIR:3308558.3313684}
Wang, Z.; Hale, S.; Adelani, D.~I.; Grabowicz, P.; Hartman, T.; Fl{\"o}ck, F.;
  and Jurgens, D.
\newblock 2019.
\newblock Demographic inference and representative population estimates from
  multilingual social media data.
\newblock In {\em The World Wide Web Conference}, WWW '19,  2056--2067.
\newblock New York, NY, USA: ACM.

\bibitem[\protect\citeauthoryear{Watts}{2007}]{watts2007twenty}
Watts, D.~J.
\newblock 2007.
\newblock A twenty-first century science.
\newblock {\em Nature} 445(7127):489.

\bibitem[\protect\citeauthoryear{Weisberg}{2009}]{weisberg2009total}
Weisberg, H.~F.
\newblock 2009.
\newblock {\em The total survey error approach: A guide to the new science of
  survey research}.
\newblock University of Chicago Press.

\bibitem[\protect\citeauthoryear{West, Sakshaug, and
  Kim}{2017}]{westAnalyticErrorImportant2017}
West, B.~T.; Sakshaug, J.~W.; and Kim, Y.
\newblock 2017.
\newblock Analytic {Error} as an {Important} {Component} of {Total} {Survey}
  {Error}.
\newblock In Biemer, P.~P.; Leeuw, E. D.~d.; Eckman, S.; Edwards, B.; Kreuter,
  F.; Lyberg, L.; Tucker, C.; and West, B.~T., eds., {\em Total survey error in
  practice}. Hoboken, New Jersey: Wiley.
\newblock  489--510.

\bibitem[\protect\citeauthoryear{Wu and Taneja}{2020}]{wu2020platform}
Wu, A.~X., and Taneja, H.
\newblock 2020.
\newblock Platform enclosure of human behavior and its measurement: Using
  behavioral trace data against platform episteme.
\newblock {\em New Media \& Society}  1461444820933547.

\bibitem[\protect\citeauthoryear{Wu \bgroup et al\mbox.\egroup
  }{2018}]{wu2018twitter}
Wu, T.; Wen, S.; Xiang, Y.; and Zhou, W.
\newblock 2018.
\newblock Twitter spam detection: Survey of new approaches and comparative
  study.
\newblock {\em Computers \& Security} 76:265--284.

\bibitem[\protect\citeauthoryear{Yang and Luo}{2017}]{yang2017tracking}
Yang, X., and Luo, J.
\newblock 2017.
\newblock Tracking illicit drug dealing and abuse on instagram using multimodal
  analysis.
\newblock {\em ACM Transactions on Intelligent Systems and Technology (TIST)}
  8(4):58.

\bibitem[\protect\citeauthoryear{Yang, Ribeiro, and
  Neville}{2017}]{yang2017should}
Yang, J.; Ribeiro, B.; and Neville, J.
\newblock 2017.
\newblock Should we be confident in peer effects estimated from social network
  crawls?
\newblock In {\em Eleventh International AAAI Conference on Web and Social
  Media}.

\bibitem[\protect\citeauthoryear{Yildiz \bgroup et al\mbox.\egroup
  }{2017}]{yildiz2017using}
Yildiz, D.; Munson, J.; Vitali, A.; Tinati, R.; and Holland, J.~A.
\newblock 2017.
\newblock Using twitter data for demographic research.
\newblock {\em Demographic Research} 37:1477--1514.

\bibitem[\protect\citeauthoryear{Yuan \bgroup et al\mbox.\egroup
  }{2013}]{yuan2013monitoring}
Yuan, Q.; Nsoesie, E.~O.; Lv, B.; Peng, G.; Chunara, R.; and Brownstein, J.~S.
\newblock 2013.
\newblock Monitoring influenza epidemics in china with search query from baidu.
\newblock {\em PloS one} 8(5):e64323.

\bibitem[\protect\citeauthoryear{Zagheni and
  Weber}{2015}]{zagheni2015demographic}
Zagheni, E., and Weber, I.
\newblock 2015.
\newblock Demographic research with non-representative internet data.
\newblock {\em International Journal of Manpower} 36(1):13--25.

\bibitem[\protect\citeauthoryear{Zagheni, Weber, and
  Gummadi}{2017}]{zagheni2017leveraging}
Zagheni, E.; Weber, I.; and Gummadi, K.
\newblock 2017.
\newblock Leveraging facebook's advertising platform to monitor stocks of
  migrants.
\newblock {\em Population and Development Review} 43(4):721--734.

\bibitem[\protect\citeauthoryear{Zhang \bgroup et al\mbox.\egroup
  }{2016}]{zhang2016your}
Zhang, J.; Hu, X.; Zhang, Y.; and Liu, H.
\newblock 2016.
\newblock Your age is no secret: Inferring microbloggers' ages via content and
  interaction analysis.
\newblock In {\em Tenth International AAAI Conference on Web and Social Media}.

\bibitem[\protect\citeauthoryear{Zhang, Hill, and
  Rothschild}{2016}]{zhang2016geolocated}
Zhang, H.; Hill, S.; and Rothschild, D.
\newblock 2016.
\newblock Geolocated twitter panels to study the impact of events.
\newblock In {\em 2016 AAAI Spring Symposium Series}.

\bibitem[\protect\citeauthoryear{Zimmer and
  Kinder-Kurlanda}{2017}]{zimmer2017internet}
Zimmer, M., and Kinder-Kurlanda, K.
\newblock 2017.
\newblock {\em Internet Research Ethics for the Social Age. New Challenges,
  Cases and Contexts}.
\newblock Peter Lang.

\end{thebibliography}

\end{document}